\documentclass[preprint,showpacs,preprintnumbers,amsmath,amssymb]{revtex4}

\usepackage{epsfig,graphicx}% Include figure files
\usepackage{dcolumn}% Align table columns on decimal point
\usepackage{bm}% bold math

\usepackage{amsmath}
\usepackage{amssymb}

\begin{document}

%\preprint{APS/123-QED}

\title{New Lagrangian diagnostics for characterizing fluid flow
  mixing}% Force
                                                               % line
                                                               % breaks
                                                               % with
                                                               % \\
% \author{R. Mundel, E. Fredj, H. Gildor,  V. Rom-Kedar }

\author{Ruty Mundel}
\affiliation{The Hebrew University
    of Jerusalem, Jerusalem 91904, Israel}

\author{Erick Fredj}
\affiliation{Department of Computer Science,  Jerusalem College of
  Technology, Jerusalem 91160,Israel}

\author{Hezi Gildor}
\affiliation{Institute of Earth Sciences, The Hebrew University
    of Jerusalem, Jerusalem 91904, Israel}

  \author{Vered Rom-Kedar} \email[]{vered.rom-kedar@weizmann.ac.il}
  \affiliation{Department of Computer Science and Applied Mathematics,
    Weizmann Institute of Science, Rehovot 76100, Israel}
  \affiliation{To whom correspondence should be addressed}

\date{\today}
\begin{abstract}
  A new kind of Lagrangian diagnostic family is proposed and a
  specific form of it is suggested for characterizing mixing: the
  maximal extent of a trajectory (MET). It enables the detection of
  coherent structures and their dynamics in two- (and potentially
  three-) dimensional unsteady flows in both bounded and open
  domains. Its computation is much easier than all other Lagrangian
  diagnostics known to us and provides new insights regarding the
  mixing properties on both short and long time scales and on both
  spatial plots and distribution diagrams. We demonstrate its
  applicability to two dimensional flows using two toy models and a
  data set of surface currents from the Mediterranean Sea.

\end{abstract}



\maketitle

%\tableofcontents

%\listoffigures

\section{Introduction}

Visualizing and quantifying mixing in unsteady fluid flows is a
magical and tricky business, with important practical implications
including larval dispersion and population connectivity
\cite{Gawarkiewicz07,bozorgmagham2013real}, oil spills
\cite{Mezic10,haller12pnas}, search and rescue
\cite{peacock2013lagrangian,melsom2012forecasting}, functioning of the
marine ecological system \cite{Dovidio09} and more
\cite{mariano2002lagrangian}. By now there are many tools to visualize
and analyze mixing properties of flows and maps
\cite{Aref02,Otti89,Mancho06,Beron-vera09,Koshel06,Peacock10,boltsan13,prants13,budivsic2012applied}. While
this field provides an endless source of scientifically produced art,
beyond its aesthetic nature lurks the scientific challenge of
characterizing these complex phenomena and providing predictions and
insights relevant for real life problems.

One aspect of the complexity arises from the flow field
structure. Unsteady flow fields typically have a mixture of Coherent
Structures (CSs), jets and mixing layers that move in an unsteady
fashion. Moreover, these structures may exist for some finite
time. Roughly, by coherent structure we mean a body of fluid which
moves together for a certain period of time, namely, we take the
Lagrangian point of view which is frame independent (see discussion
and references in \cite{boltsan13,PoHa99,hallerjfm13}). Passive
particles placed inside such a coherent structure remain in it as long
as it lives, moving roughly quasi-periodically around the coherent
structure center. Here we mainly focus on such CSs. Jets may be
similarly characterized as regular particles that flow between neighboring
sections.  These structures are typically separated by mixing layers,
the regions in which there is ``chaotic mixing" - a sensitive
dependence of the Lagrangian trajectories on initial conditions
(i.c.). Particles belonging to the mixing layer may stick to a nearby
coherent structure or a jet\ for a certain period and then eject from
it. This complex mixture of structures may appear in flows in closed
domains (such as closed basins), in open domains (such as coastal
areas) or in practically unbounded domains (such as eddies within the
Pacific Ocean).

Another aspect of the complexity is the infinite dimensional nature of
the initial data problem
\cite{pier94,lin2011optimal,budivsic2012applied}. Indeed, the initial
distribution of the particle density belongs to the space of all
possible initial distributions of scalar fields. Different mixing
characteristics may apply to particular subclasses of such
distributions \cite{LiBe95,mathew2005multiscale}.

The last aspect we mention here is the temporal complexity of the
problem. There are the classical mixing time scales associated with
the molecular diffusion and viscosity, relevant for both steady and
unsteady flows. However, for unsteady flows, additional scales, those
associated with the unsteady component frequencies and amplitudes and
those associated with the resulting chaotic mixing scales, emerge
\cite{BeLeWi91,RkPo99}. Finally, in many applications the observation
time scale is also relevant \cite{Haller00,PoHa99}.

Defining a proper characterization of mixing is non-trivial and is
problem and application  oriented. Indeed, with all these complexities
in mind, with the common appearance of mixture of flow regimes having
temporal variations, we conclude that any classification scheme of
mixing domains must have some tunable threshold parameters. This observation
impedes the quest for objective classification.  Indeed, despite being
a classical long standing problem, new mixing characteristics are
suggested and presented in various ways, from both Eulerian and
Lagrangian points of view
\cite{Otti89,boltsan13,prants13,Shadden05,Orre06,Boffetta01,lipphardt2006synoptic}.
% hezi: \cite{Otti89,boltsan13,prants1,Boffetta013,Shadden05,Orre06}
% missing Boffetta.

Eulerian characteristics correspond to snapshots (or temporal
averages) of the velocity field or its spatial derivatives (e.g. the
Okubo-Weiss criterion or the vorticity field
\cite{okubo70,weiss91}). In contrast, Lagrangian characteristics are
based on an integrative procedure by which observables are measured
along trajectories (e.g. the absolute dispersion (AD) measures the
distance travelled by a particle, the relative dispersion (RD)
measures the distance between a particle and its neighbors, the
finite-time Lyapunov exponent (FTLE) measures the maximal local
stretching rate etc.). Some of the Lagrangian characteristics present
the resulting observable after a certain integration time, with no
information regarding the intermediate time dynamics (e.g. the AD and
RD fields depend only on the initial and final location of the
particles), whereas some of the other Lagrangian characteristics use
averaging or integration along the trajectories
\cite{Haller00,Mezic10,Ryp12,MendozaManco10,Alvaroet1013,budivsic2012applied}. These
Lagrangian fields are commonly used to identify regions of small and
enhanced stretching and, in particular, are used to identify the
spatial position of dividing surfaces between different regions, the
Lagrangian Coherent Structures (LCS)
\cite{Haller00,Shadden05,bozorgmagham2013real}. Another approach,
mainly applicable for time-dependent open flow is based on the
residence time that particles spend in a certain domain
\cite{lipphardt2006synoptic,RoLW90}.  The locations and size of CSs has
been mainly studied by the transfer-operator approach, providing a
connection between the Eulerian and Lagrangian perspectives
\cite{dellnitz1999approximation,froy07,Fropad12,boltsan13}. More
recently, the notion of coherent Lagrangian vortices was introduced to
identify CSs by using a variational principle on the averaged
Lagrangian strain \cite{hallerjfm13}.  In \cite{thesis08} the notion
of maximal absolute dispersion was introduced for studying the lobe
dynamics for surface particles embedded in a three dimensional
velocity field \cite{AhGilRK12}. This study has motivated much of the
current work.

Here we propose a new family of Lagrangian characteristics: the
spatial dependence of extreme values of an observable along
trajectories. Since asymptotically this value in each ergodic
component converges to a common extreme value (similarly to other
Lagrangian averages along trajectories
\cite{budivsic2012applied,Mezic10,Ryp12}), such extremal fields
provide the sought division into distinct ergodic components.
%\footnote{This
%  theoretically appealing property, aside of the general context and
%  the directional addition here, is the main difference between this
%  notion and the motivating study of the maximal absolute dispersion
%  initiated in \cite{thesis08}}.
Moreover, the extreme values of the selected observable may by
themselves have significant physical meaning. Examples of such
significant observable values are: maximal/minimal locations of the
particle in a certain direction (hereafter MET), maximal speed or
strain experienced by the particle, closest approach to the particle
initial location or closest approach to a prescribed location. In
fact, any of the commonly calculated Lagrangian fields may be chosen
as an observable.

Here we focus on the MET, the extreme location of particles in a
certain direction.  These new characteristics have a few
advantages. First, their computation cost is relatively low. Second,
by definition their convergence in time is non-oscillatory. Third, and
most important, by choosing the MET and examining its Cumulative
Distribution Function (CDF) we can extract not only the existence of
CSs, but can also quickly determine many of their characteristics
(e.g., their number and size). This feature will potentially allow for
a substantial data reduction; see below.

Studying extreme value statistics in the context of chaotic dynamical
systems is a fascinating relatively new field of research
\cite{Collet2001,Holland2012,Lucarini2012,Freitas2008,Freitas2013}.
Previous works on the extreme values of an observable of dynamical
systems have focused on the temporal dependence of a single chaotic
trajectory for maps (mainly for chaotic dissipative maps), connecting
it to the universal distributions appearing in the field of Extreme
Value Statistics (EVS) on one hand and to Poincar\'e recurrences and
local dimensionality of the attractors on the other (the connection to
\cite{Ryp12,Lucarini2012} may thus be intriguing).  Here, we focus
instead on utilizing the extreme value functionals as convenient
spatial characteristics of dynamical systems with mixed phase space.

The paper is ordered as follows: we first define the new family of
characteristics (section II) and explore their properties using a few
toy models (section III). We then apply these measures to real
geophysical data - surface currents in the eastern Mediterranean
(section IV). We conclude and discuss some of the future directions in
section V.

\section{main concepts and definitions}

Consider the motion of passive particles in a fluid flow (see specific
examples below):\begin{equation} \frac{dx}{dt}=u(x,t),\qquad\
  x\in\mathbb{R}^n, \ n=2\text{ or } 3
\end{equation}and consider the extremal values of an observable
function \(\phi\) for each particle along a time segment of a trajectory:
\begin{eqnarray}
  M^{+}_{\phi}(\tau ;x_{0},t_{1})&=\max_{t\in[t_{1},t_{1}+\tau]} \phi (x(t;t_{0}))  \\
  M^{-}_{\phi}(\tau ;x_{0},t_{1})&=\min_{t\in[t_{1},t_1+\tau]} \phi (x(t;t_{0}))
  \\
  M_{\phi}(\tau ;x_{0},t_{1})&=M^{+}_{\phi}(\tau ;x_{0},t_{1})-M^{-}_{\phi}(\tau ;x_{0},t_{1})
\end{eqnarray}
where \(x(t_{0};t_{0})=x_{0}\) and \(t_1\in\mathbb{R}\). Notice there
are three time parameters in the above definition: \(t_{0}\)
corresponds to the seeding time of the particles - the velocity field
phase at which the integration of the trajectories
begins. \([t_{1},t_{1}+\tau]\) is the extremal window, the recording
time interval on which the observable is maximized/minimized. One
natural choice is to take \(t_{1}=t_0\) and \(\tau\) sufficiently
large with respect to the CS turnover time, so that the CS is resolved
within the extremal window. For periodic flows, shifting the extremal
window may reveal trapping regions of the CSs. For unsteady flows,
when coherent structures emerge and disappear or move around in an
unknown manner, windowing in \(t_{1}\) and \(\tau\) may reveal the
temporal existence and spatial movement of CSs, see below.
Asymptotically, we define
\begin{eqnarray}
  M^{+}_{\phi}( x_{0})&=\limsup_{t} \phi (x(t;t_{0}))  \\
  M^{-}_{\phi}( x_{0})&=\liminf_{t} \phi (x(t;t_{0}))
  \\
  M_{\phi}(x_{0})&=M^{+}_{\phi}(x_{0})-M^{-}_{\phi}(x_{0})
\end{eqnarray}
with similar definitions for the negative time asymptotic. Notice that
$M_{\phi}(0;x_{0},t_{0})={0,\,\ }M^{+}_{\phi}(0
;x_{0},t_{1})=M^{-}_{\phi}(0 ;x_{0},t_{1})=\phi
(x(t_{1};t_{0}))\)\($, and, $M_{\phi}(\tau
;x_{0},t_{1}),M^{+}_{\phi}(\tau ;x_{0},t_{1}),-M^{-}_{\phi}(\tau
;x_{0},t_{1})$ are nondecreasing functions of $\tau$.

These definitions naturally extend to maps. Indeed, for time-periodic
flows, when the observation time includes many periods, it makes sense
to consider the discrete time series found from the Poincare map (the
stroboscopic sampling of the signal) instead of the continuous flow,
and all the above notions apply. Here, however, for deductive reasons
we do not use the time-periodicity feature of the toy models. Instead,
we keep in mind the general setting for geophysical flows where the
velocity field is not periodic and in principle even when there is a
known dominant frequency in the spectra the observation time may be
shorter than the associated period.

Here we focus on the MET by setting the observable \(\phi\) to measure
the extent of the particle position in a given direction
\(\mathbf{r}\): \(\phi(x(t;t_{0}))=x(t;t_{0})\cdot \mathbf{r}\).
\(M^{\pm}_{\bf r}(\tau ;x_{0},t_{1}) \) represent the maximal/minimal
extents visited by the particle during the extremal window
\([t_{1},t_{1}+\tau]\).  \(M^{}_{\bf r}(\tau ;x_{0},t_{1}) \) denotes
the difference between the maximal and minimal extents, and is called
the \textbf{maximal shift}.

\begin{figure}
\centering
\includegraphics[width=12cm]{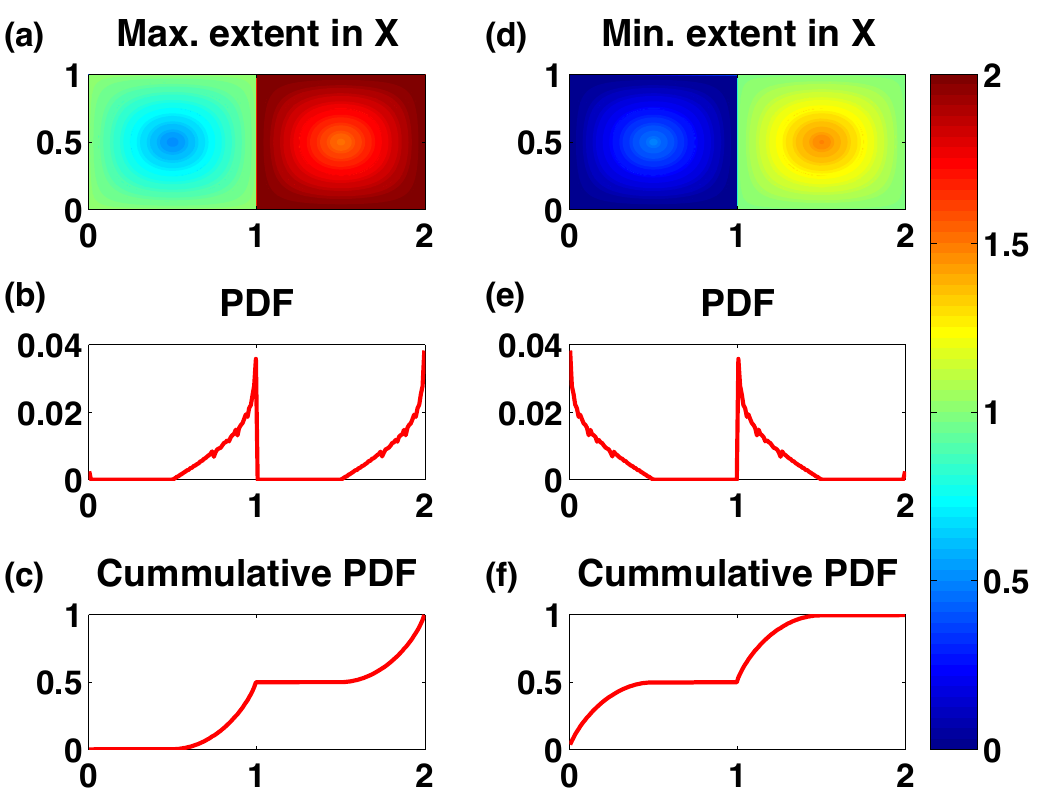}
\caption{Two stationary coherent structures: the maximal (left) and
  minimal (right) \(x \) position (\( M^{\pm}_{\bf (1,0)}\)) of all
  initial conditions is plotted (top), as are the corresponding PDF
  (middle) and CDF (bottom) functions. Each gyre appears in the PDF
  and correspondingly in the CDF: the left gyre accumulates to occupy
  exactly half of the domain, as apparent from the plateau region. The
  centers and extents of the two gyres are clearly
  seen. Eq.~\ref{eq:xyPerturb}, \(A=0.25, \epsilon=0\) and
  \(\tau=200\).
  %RUTY: STILL\ NEED THE\XMIN plot�\
}
\label{fig:singlecs}
\end{figure}

We propose that by monitoring these fields, which are trivial to
compute, we can infer quite a few properties of the Lagrangian flow
structure both asymptotically and transiently.  Moreover, we propose
that such properties may be found quite efficiently by analyzing the
cumulative distribution function (CDF) of the MET field. This may lead
to significant data reduction from 2D fields maps (e.g. of the FTLE or
RD) to a 1D plot. Often, especially in realistic geophysical
applications, the amount of data (e.g. data extracted from satellites
or from general circulation models) is huge and time-dependent, and
efficient data-compression methods are needed. As described in the
next section, the shape of the CDF provides information about the
existence of coherent structures, their locations, and the existence
of chaotic zones. It seems that some of the properties may even be
inferred from a sampling of the flow field in only a few directions,
making this diagnostic a potentially useful tool in real applications
allowing a limited sampling of the flow (e.g., by only few drifters)
even in the fully three dimensional setting. We will further explore
this direction in future studies.

%one-dimensional sampling of the flow field and this sampling may be
%relevant possibly also in 3D - this direction will be developed in
%later publications.

\section{Typical features of the MET-Toy models}

Next we examine the properties of the extremal fields at typical
structures that appear in unsteady flows. To this aim we first
consider prototypical models for stirring and mixing in bounded
domains - the steady and time-dependent double gyre models. We then
consider the oscillating vortex pair model to demonstrate the method
on a flow in an unbounded domain.

A CS, loosely defined as a group of trajectories with a common
averaged behavior on some fast eddy-turnover time scale, may be
stationary, gently oscillating, rotating or advected in an unbounded
domain. The FTLE field for particles in all such structures
asymptotically vanishes. Below, we list the characteristic features of
the MET in these different settings. We show that while the MET is
simpler to compute it provides additional information about the
properties of the coherent structures.  To gain intuition we examine
the double gyre model \cite{Shadden05}:
\begin{eqnarray}
  \Psi_{xy}(x,y,t)&=&  A \sin(\pi f(x,t))\sin \pi y   \label{eq:xyPerturb}
  \\
  f \left(x,t\right) &=& \epsilon \sin\omega t\ x^2+(1-2\epsilon \sin\omega t)x.\  \nonumber
\end{eqnarray}
We begin with the trivial case of the steady double gyre
(\(\epsilon=0)\) and then continue to more realistic settings.

\textbf{A single stationary coherent structure. } Consider the
stationary double gyre model.  The maximal and minimal extents in the
\(x\) direction for this case are shown in
Fig.~\ref{fig:singlecs}a,d. The flow has two symmetric gyres lying
along the horizontal direction and no mixing zone. Examine the left
gyre first.  All trajectories belonging to the left gyre are bounded
and periodic in time. Hence, for any fixed \(t_{1}\), for all
\(\tau\), \(M_{\bf r}(\tau ;x_{0},t_{1})\) and \(M^{\pm}_{\bf
  r}(\cdot)\) are finite and, for sufficiently large \(\tau\),
\(M_{\bf r}(\cdot)\) is equal to the width of the periodic orbit in
the direction \(\mathbf{r}\), whereas \(M^{\pm}_{\bf r}\) provide the
maximal and minimal extents in this direction.

{\bf Asymptotic form of the PDF and CDF } The PDFs and CDFs for this
case are also shown in Fig.~\ref{fig:singlecs}. The CDF of \(M^+_{\bf
  r}\) converges to a piecewise smooth increasing function which
starts increasing quadratically from zero concentration at the
coherent structure center \((x=0.5) \) and abruptly stops increasing
at the coherent structure boundary (\(x=1.0)\). The CDF of \(M^-_{\bf
  r}\) (Fig.~\ref{fig:singlecs}f) starts increasing abruptly at the
coherent structure leftmost boundary \((x=0) \) and stops increasing
quadratically at the coherent structure center (\(x=0.5)\).  The area
fraction of the left coherent structure is the CDF value at the
plateau.

\begin{figure}
 \centering \includegraphics[width=15cm]{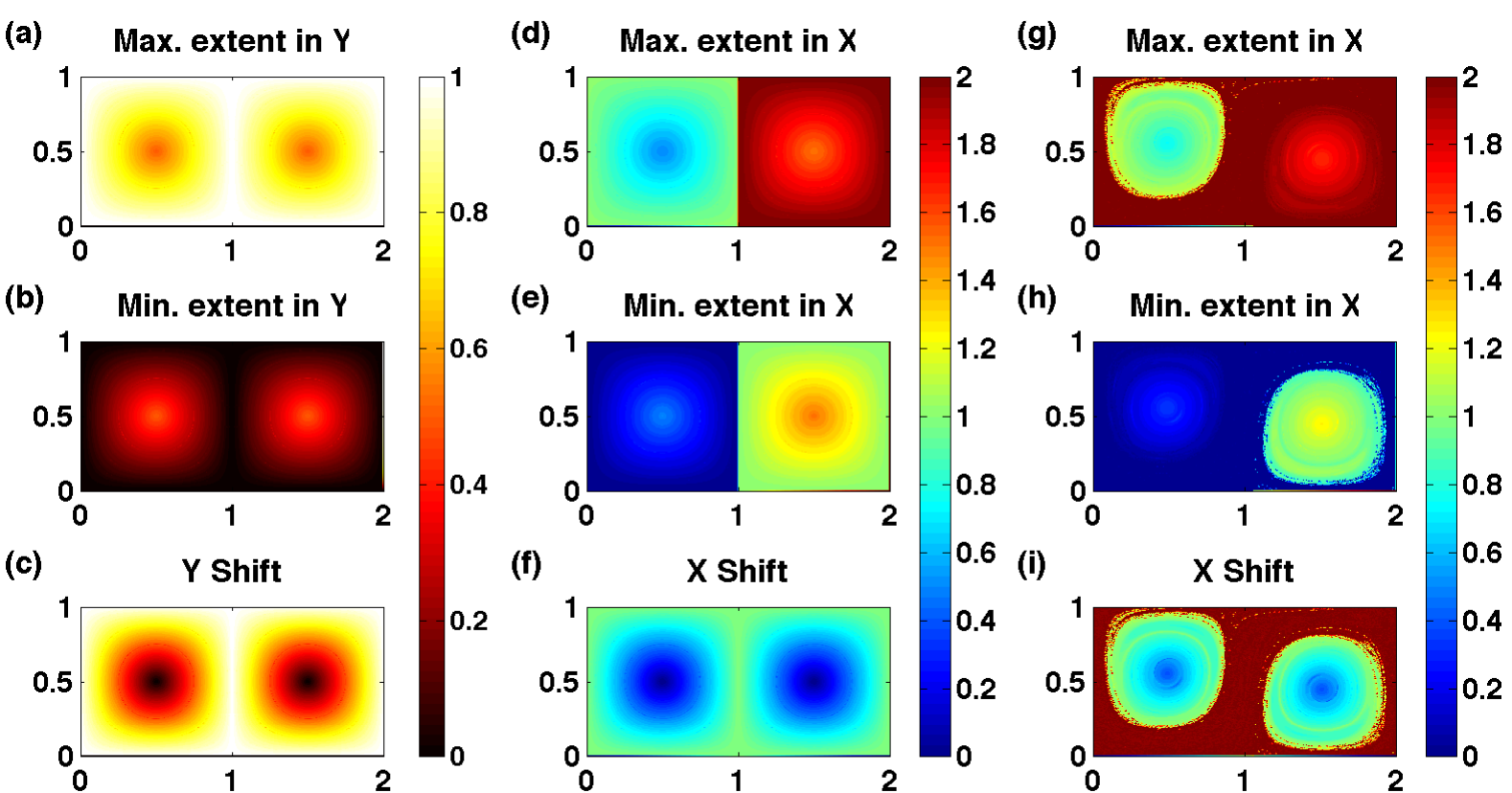}
 \caption{Two stationary (a-f) and oscillatory (g-i) coherent
   structures.  a-c) The need to choose resolving direction for
   obtaining the number of distinct CSs from the corresponding CDFs is
   demonstrated. The two gyres here have exactly the same extent in
   the \(y\) direction (i.e. identical \( M^{\pm}_{\bf (0,1)}\)
   values) hence the CDF of the \( M^{\pm}_{\bf (0,1)}\) does not lead
   to a distinction between the left and right gyres. d-f) The need
   for the maximal/minimal quantifiers is demonstrated - while the
   maximal/minimal \(x \) values (\( M^{\pm}_{\bf (1,0)}\))
   distinguish between the right and left gyres, their difference, the
   maximal shift \( M^{}_{\bf (1,0)}\), does not. g-i) The
   applicability to unsteady flow is demonstrated
   (Eq.~\ref{eq:xyPerturb} with \(A=\epsilon=0.25, \omega=2\pi/10\)
   and \(\tau=200\)). Notice that in both the stationary and
   oscillatory cases, at the coherent structure centers the fields
   \(M^{}_\mathbf{r},M^{+}_\mathbf{r},-M^{-}_\mathbf{r}\) attain their
   local minima. In the stationary case (a-f) the value of
   \(M^{\pm}_\mathbf{r}\) at the center matches the center position
   whereas in the oscillatory case there is a mismatch due to the
   oscillation:
   \(M^{\pm}_\mathbf{r}-x_c(t_0;t_0)\cdot\mathbf{r}\neq0.\) }
\label{fig:severalcs}
\end{figure}

 \textbf{Several stationary coherent structures. }
%EPS=0 a=1
 The need for the three quantifiers \(M_{\bf r}, M^{\pm}_{\bf r}\) and
 the directional dependence is clarified when several coherent
 structures coexist in the flow. Fig.~\ref{fig:severalcs} shows these
 three fields for two directions (\(x\) and \(y\)). First, we observe
 that the \(M_\mathbf{r}^\pm\) fields distinguish between different
 CSs provided that the structures have no overlap in the direction
 \(\mathbf{r}\). In our example it is clear that the \(x\)
 maximal/minimal extent fields (\(M_\mathbf{(1,0)}^\pm\),
 Figs.~\ref{fig:singlecs} and \ref{fig:severalcs}d,e) distinguish
 between the left and right gyres whereas the \(y\) maximal/minimal
 extent fields (\(M_\mathbf{(0,1)}^\pm\), Fig.~\ref{fig:severalcs}g,h)
 do not. More generally, denote by \(L_{x-left}\) the \(x\) coordinate
 of the right boundary of the left gyre, by
 \((C_{x-right},C_{y-right})\) the center of the right gyre and by
 \(L_{y}\) the height of the gyres. Then, for
 \(\mathbf{r}=(r_{x},r_y)\), we see that \(M_\mathbf{r}^+
 (\text{i.c. in left gyre} )=\max_\text{i.c. in left
   gyre}(x(t)r_x+y(t)r_y)<L_{x-left}r_x+L_yr_y\) whereas
 \(M_\mathbf{r}^+(\text{i.c. right gyre})=\max_\text{i.c. in right
   gyre}(x(t)r_x+yr_y)>C_{x-right}r_x+C_{y-right}r_y\). Hence, if
 there is a gap between the two bounds (here
 \(\frac{r_y}{r_x}<\frac{C_{x-right}-L_{x-left}}{L_y-C_{y-right}}\) )
 we will call \textbf{r} a \textbf{resolving direction} and then the
 CDFs of \(M_\mathbf{r}^\pm\) show two distinct monotone increasing
 regimes each corresponding to a different gyre as in
 Fig.~\ref{fig:singlecs}.  On the other hand, we notice that the
 \(M_{\bf r}\) field is identical for the two gyres, and more
 generally, for all \textbf{r}, all the coherent structures are lumped
 together in this field (similarly to the FTLE and RD fields).

 More generally, we see that depending on the structures' alignments,
 a direction \(\mathbf{r}\) may or may not resolve the structures. If
 the center of one structure is bounded away from the maximal (or
 minimal) extent of the other in the direction \(\mathbf{r}\) we do
 have separation - a gap in the \(M_\mathbf{r}^\pm\) values. We expect
 to be able to find such resolving directions when there is a small
 number of coherent structures, but not when there are many possibly
 disordered structures in the domain (as in 2D-turbulent flows), see
 discussion.

 The important conclusion from the above is that the CDF of the
 \(M_\mathbf{r}^\pm\) fields with a resolving direction \textbf{r} may
 be used to distinguish between the existence of multiple vs. a single
 CSs (thus helping in data reduction). In contrast, the CDF of the
 \(M_\mathbf{r}\) field, of the MET in non resolving directions, of
 the FTLE field and of other similar fields cannot help in counting
 the number of distinct CSs.

\begin{figure}
\centering
\includegraphics[width=5cm]{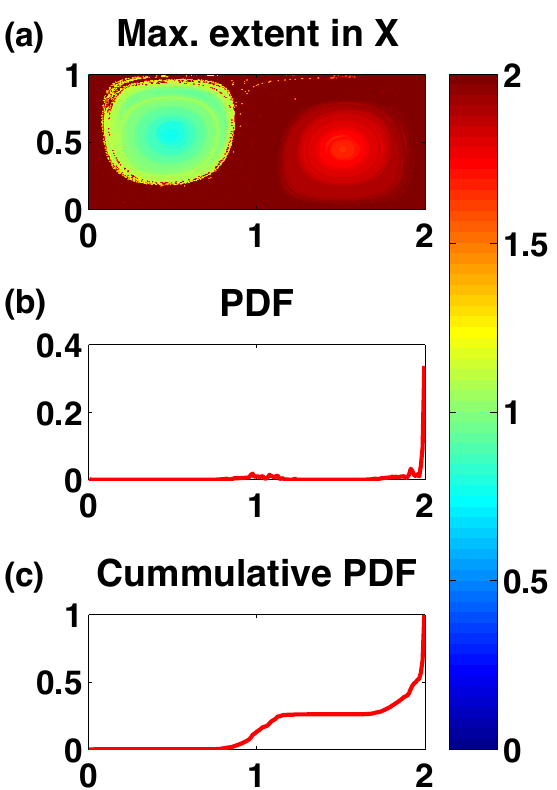}
\caption{Oscillatory double gyre - the maximal extent in \(x\), \(
  M^{+}_{\bf (1,0)}\), its PDF, CDF and the mixing layer.  The two
  gyres and the mixing layer are clearly distinguishable in the CDF,
  whereas in the PDF the high concentration of the distribution
  function in the mixing layer bin (\(x_{max}\lesssim2\)) makes the
  contributions to the regular coherent structures nonvisible.
  \(A=\epsilon=0.25, \omega=2\pi/10\) and \(\tau=200\). }
\label{fig:pdfcdf}
\end{figure}

\begin{figure}
\centering {
  \includegraphics[width=7.5cm]{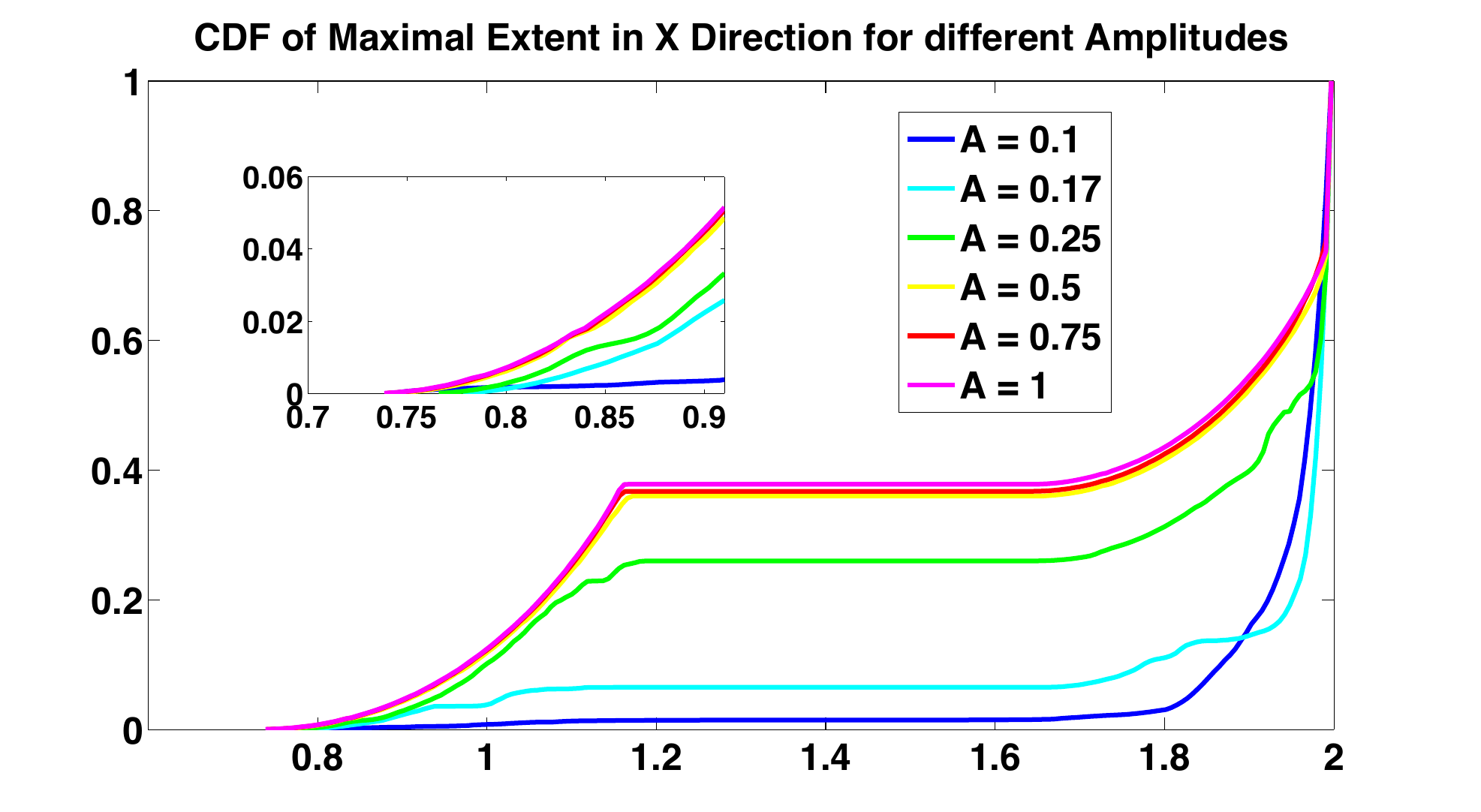}
  \includegraphics[width=8.5cm]{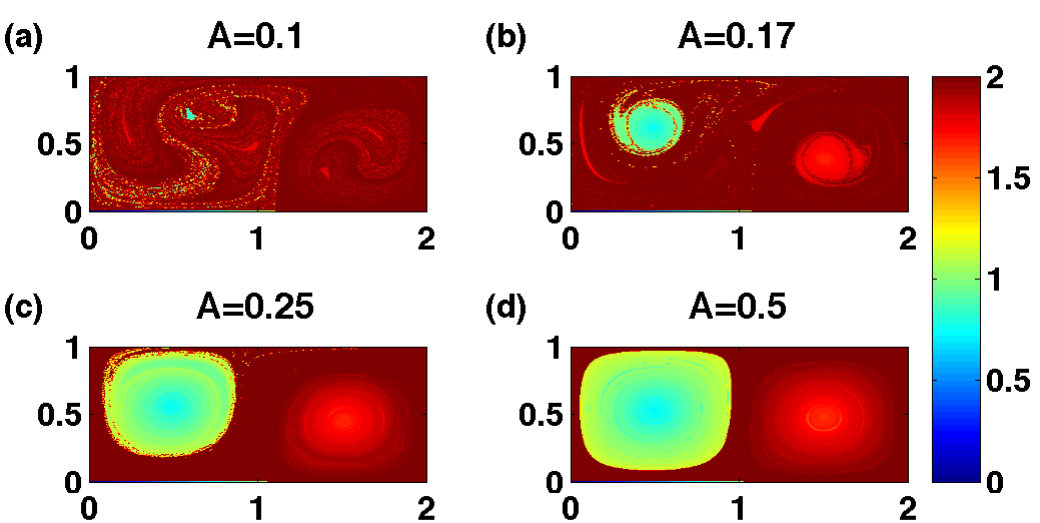}
      }
      \caption{Oscillatory double gyre for different base flow
        intensity $A$ after 20 periods. Left: The CDF provides
        succinct presentation of the CS size and location for the
        different flows without using any flow visualization analysis.
        Right: The maximal extent in \(x\), \( M^{+}_{\bf (1,0)}\), is
        shown for four different base flow strengths.  The two gyres
        and the mixing layer are clearly distinguishable. The
        difference between the location of the gyre center and its
        maximal value provides information regarding its oscillation
        magnitude.  The accumulated size of the coherent structures is
        seen to decrease here with decreasing \(A\) thus providing
        estimates for the area of the mixing zone.  \(\epsilon=0.25,
        \omega=2\pi/10\) and \(\tau=200\).  }
\label{fig:oscilcs}
\end{figure}

\textbf{Oscillating coherent structures and the mixing layer }
Consider now the periodically perturbed double gyre model shown in
Fig.~\ref{fig:severalcs}g-i.  Here, trajectories starting inside the
two gyres move on some invariant rings around the two oscillating
centers located near \(x=0.5,1.5,y=0.5\). We expect that most of the
initial conditions in these gyres belong to KAM tori, namely they
perform regular (quasiperiodic) motion. We observe that the properties
of the fields \(M^{}_\mathbf{r},M^{\pm}_\mathbf{r}\) and their CDFs in
the CSs are similar to those of the steady gyres
(Fig.~\ref{fig:pdfcdf}). To gain intuition, assume that the
trajectories are of the form: \(x(t;t_{0})=x_{c}(t;t_{0})+g(t;x_{0})
\) where \(x_{c}(t;t_{0})\) is some unknown slowly moving center\ and
\(g\) is rapidly oscillating with zero mean (otherwise the particle
drifts away from the center). The main difference between these and
the stationary CSs, and in fact a way to identify these oscillating
structures, appears when one examines the value of
\(M_\mathbf{r}^\pm\) at the coherent structure center
\(x_{c}(t;t_{0})\), where, as before, we may define
\(x_{c}(t_0;t_{0})\) as the trajectory along which \(M_\mathbf{r}\)
attains its local minima. In the steady case, we have
\(M_\mathbf{r}^-=M_\mathbf{r}^+=x_{c}(t_0;t_{0})\cdot \mathbf{r}\),
whereas in the oscillating case these values provide the minimal and
maximal central location of the coherent structure along the
\(\mathbf{r}\) direction, see Fig.~\ref{fig:severalcs}g-i (see also
Fig.~\ref{fig:oscilcs}).

In the oscillatory case a mixing layer appears: it consists of chaotic
trajectories having sensitive dependence on i.c. that eventually
encircle both gyres.  Hence, for all i.c.'s \(x_0\) in the mixing
layer, the values of \(M_{\bf r}, M^{\pm}_{\bf r}\) asymptote to the
width/the extent of the mixing layer in the \(\mathbf{r}\)
direction. Hence, the PDFs of \(M_{\bf r}, M^{\pm}_{\bf r}\) converge
to a delta function on the chaotic component, at the value of the
maximal extent of the mixing layer (in the present case the maximal
extent of the domain) - the chaotic bin. In the PDF of these fields
the only observable structure is the mixing layer whereas the regular
coherent structures become invisible (Fig.~\ref{fig:pdfcdf}b). In the
CDF plot the finite volume of the chaotic layer is apparent
(Fig.~\ref{fig:pdfcdf}c).

\begin{figure}
\centering {
  \includegraphics[width=7.5cm]{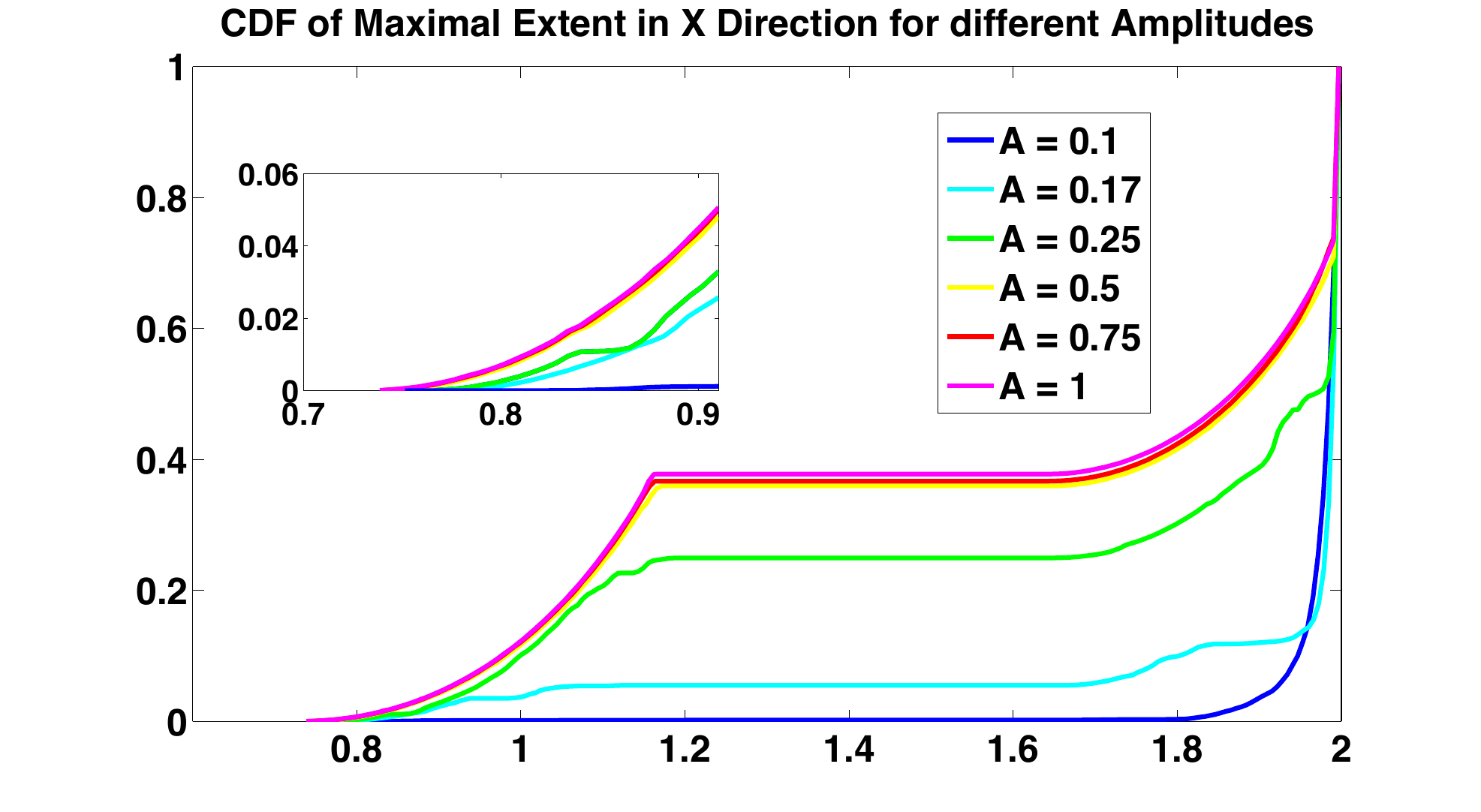}
  \includegraphics[width=8.5cm]{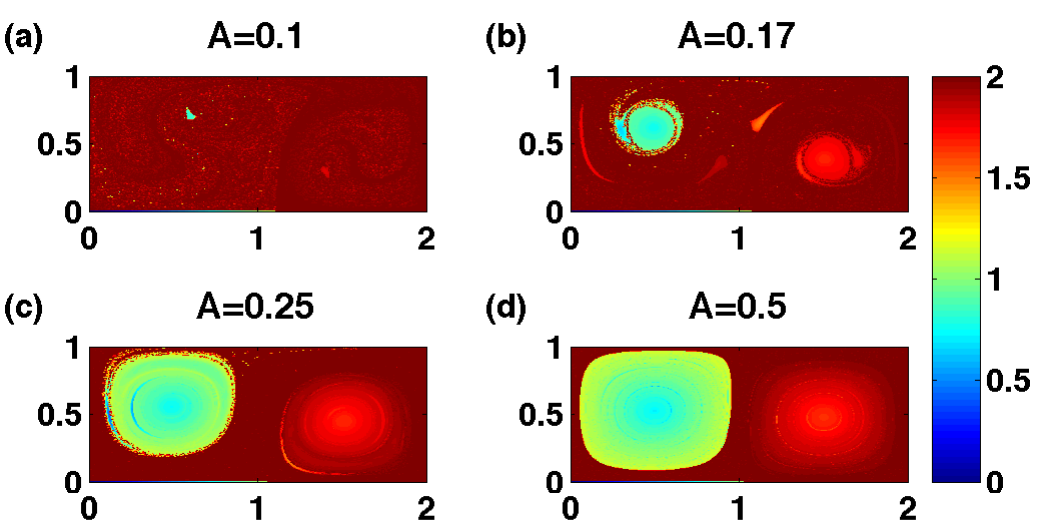}
 }
 \caption{Oscillatory double gyre for different base flow intensity
   \(A,\) as in Fig \ref{fig:oscilcs} for double the integration time
   (40 periods). The convergence of the CDFs and of the MET fields in
   regular regions is mainly achieved already after 20 periods,
   whereas the mixing layers have slower convergence and do not
   achieve their asymptotic behavior even after 40 periods.
   \(\epsilon=0.25, \omega=2\pi/10\) and \(\tau=400\).}
\label{fig:oscilcslong}
\end{figure}

The boundary between the mixing layer and the coherent structure is
especially interesting - the MET are discontinuous at this boundary
(Fig.~\ref{fig:pdfcdf},\ref{fig:oscilcs},\ref{fig:oscilcslong}).
Moreover, the convergence characteristics of these fields are
different in the mixing vs. the CS regions. High variability is
expected in the chaotic zone whereas in the coherent structures the
convergence is expected to be regular.

Finally, Figs.~\ref{fig:oscilcs} and \ref{fig:oscilcslong} show that
the CDFs of the MET fields provide a succinct way to compare the
mixing properties of different flow fields. In these figures we
compare the CDFs of the unsteady double gyre model for decreasing
power of the gyre intensity after 20 (Fig.~\ref{fig:oscilcs}) and 40
(Fig.~\ref{fig:oscilcslong}) periods. By decreasing the gyre intensity
(\(A\) in Eq.~\ref{eq:xyPerturb}) we effectively increase the
non-dimensional period of the oscillatory component \cite{RkPo99}.
The CDFs reveal how the two CSs centers oscillate to larger extent and
shrink in size with \(A\) without using any flow visualization
analysis. Indeed, notice that the parabolic increase in the CDF
starting near \(x=0.7\) (\(x=1.7)\) corresponds to the maximal
\(x\)-locations of the left (right) CS center respectively. Thus, the
change in this value with \(A\) (see insert) indicates that the
centers experience larger oscillations as \(A\) decreases. The plateau
value of the CDF provides the area of the left CS (seen to decrease
from 0.4 to nearly vanishing values as \(A\) decreases). The sharp
increase in the CDF towards \(x=2.0 \) indicates the transition to the
mixing layer orbits, thus, for sufficiently large \(\tau\), the value
of the CDF at the transition point provides the total area of the
regular component. Notice that when \(A\) decreases to \(0.17\) the
two gyres break into smaller CSs, some of these begin to rotate in the
box (see the bright red crescent on the left part of the box - this
crescent together with the island to the right of the center line
correspond to a period two CS. The large discrepancy between the
location of this crescent and the $M^{+}_{\bf (1,0)}$ value in it,
which is equal to that found in the other island suggests its
rotational nature. This may be verified by trajectory computations and
by looking at the minimal extent field, not shown). The above
observations apply to a sufficiently long extremal window
$\tau$. Comparing Figs.~\ref{fig:oscilcs} and \ref{fig:oscilcslong} it
is seen that while the CDF component of the CSs appears to converge
already after 20 periods the mixing component has not reached its
asymptotic form even after 40 cycles.  Indeed, the ghost of the
stable manifold is readily seen for short extremal windows. These
distinct convergence properties and the transient features of the MET
may be utilized to distinguish between different regions and for
locating dividing surfaces - this is left to future studies.

\textbf{CS in unbounded flows} Next we consider an open flow
  model, the Oscillating Vortex Pair (OVP): a vortex pair in an
  oscillating strain-rate field embedded in a uniform flow.  The
  non-dimensional stream function \cite{RoLW90} is of the form
  \footnote{Remark: in the dimensional units (see \cite{RoLW90}
    eq.~2.1-2.4): \(v=\frac{2\pi dV_{v}}{\Gamma},
    \epsilon=\frac{2\pi\varepsilon d^{2}}{\Gamma},
    \omega=\frac{2\pi\varpi d^{2}}{\Gamma} \) where \(d,\Gamma
    ,V_{v},\varepsilon,\varpi\) denote, respectively, half the initial
    distance between the vortices, their strengths, the moving frame
    velocity, the strength of the strain field and its frequency. In
    particular \(\epsilon\) here is \(\epsilon/\gamma\) in eq.~2.3-2.4
    of \cite{RoLW90} and \(\omega\) here is \(1/\gamma\) there.  }:
\begin{eqnarray}
\nonumber
  \psi(x,y,t)&=& -\log\frac{(x-x_{v})^2+(y-y_{v})^2}{(x-x_{v})^2+(y+y_{v})^2}
  - vy +\epsilon xy\sin(\omega t) \\
\frac{dx_v}{dt}&=&\frac{1}{2y_{v}}-v+\epsilon
  x_{v}\sin(\omega t),\label{eq:psivortexpair}\\
  \nonumber
  \frac{dy_v}{dt}&=&-\epsilon
  y_{v}\sin(\omega t)
  \end{eqnarray}
  where \((x_{v}(t),\pm y_v(t)) \) denotes the vortex locations and
  \(x_v(0)=0, y_v(0)=1\). The vortex pair moves, in an oscillatory
  fashion, in the positive \(x\) direction with an average horizontal
  velocity \(v_{vort}=0.5-v+O(\epsilon) \). As the vortex pair advects
  it carries with it a body of fluid, and due to the oscillations it
  sheds parts of this body of fluid in the form of lobes, see
  \cite{RoLW90}. There, \(v\) was tuned so that \(v_{vort}\approx 0\)
  and thus the mixing was observed in the vortex pair moving
  frame. Here we take different values of \(v\) to examine the
  dependence of the MET methodology on the frame of reference \footnote{To speed up the numerical
    computations of the passive particles that are placed very close
    to the vortices we replace the small denominators in the velocity
    field by a cut off value of 0.01.}.

\begin{figure}
\centering
\includegraphics[width=5cm]{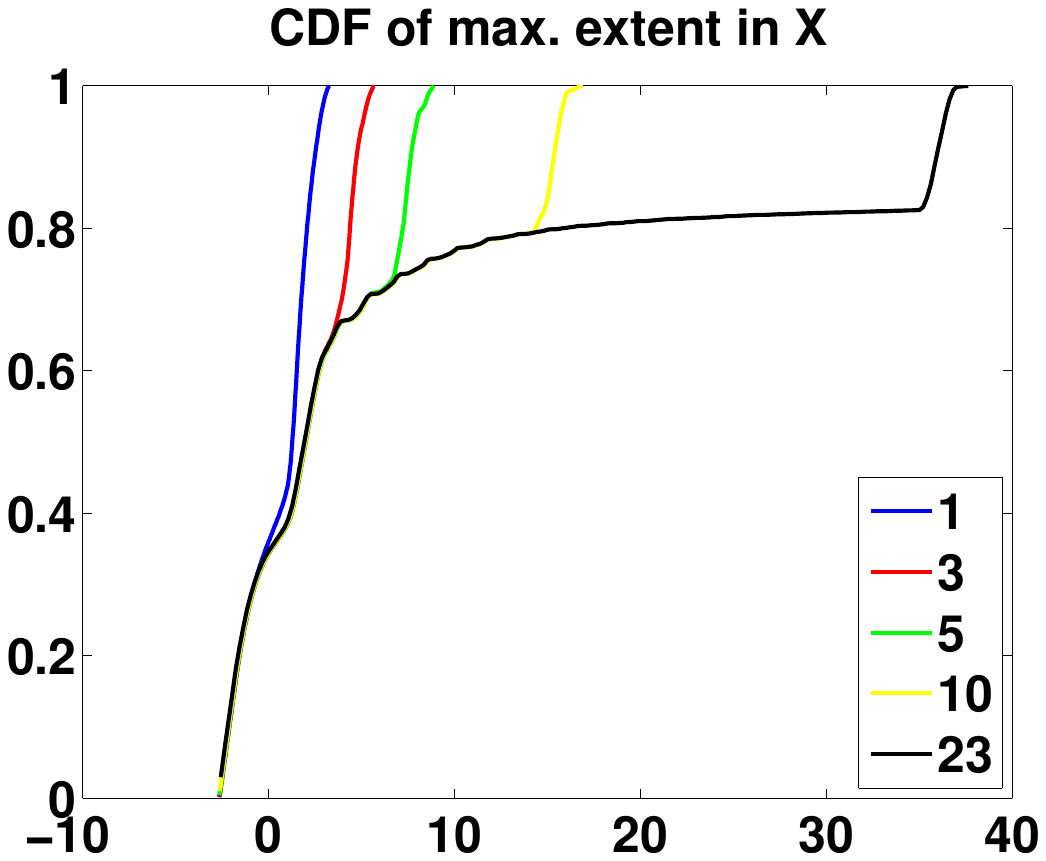}
\includegraphics[width=7cm]{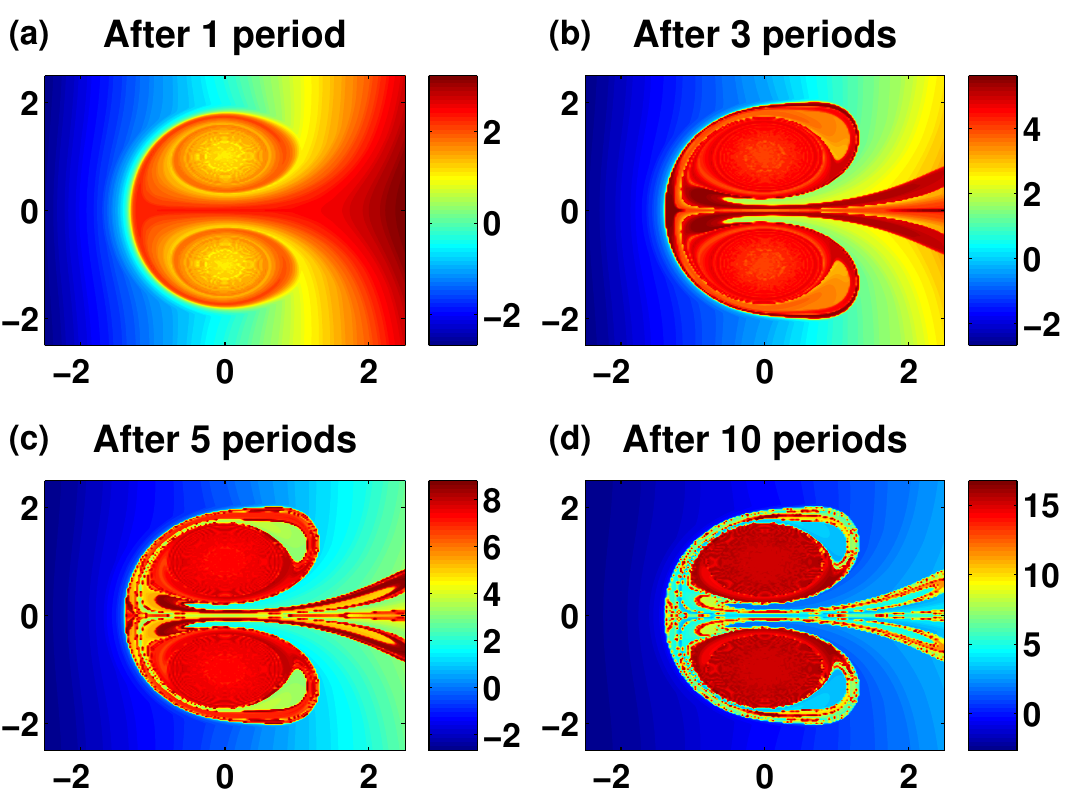}
\caption{The MET for the OVP open flow - dependence on time. The four
  right panels show the maximal extent in $x$, $M^{+}_{\bf (1,0)}$, at
  four different extremal windows ($\tau=(1,3,5,10)\cdot2\pi/\omega$)
  for the OVP flow ( Eq. \ref{eq:psivortexpair} with
  $\epsilon=0.2, \omega=1.45, v=0.25$). The left panel shows the CDF
  of this flow at the corresponding times - the rightmost curve after
  23 periods. The location and size of the vortical core may be easily
  identified from the CDF. }
\label{fig:movingtime}
\end{figure}

\begin{figure}
\centering
\includegraphics[width=5cm]{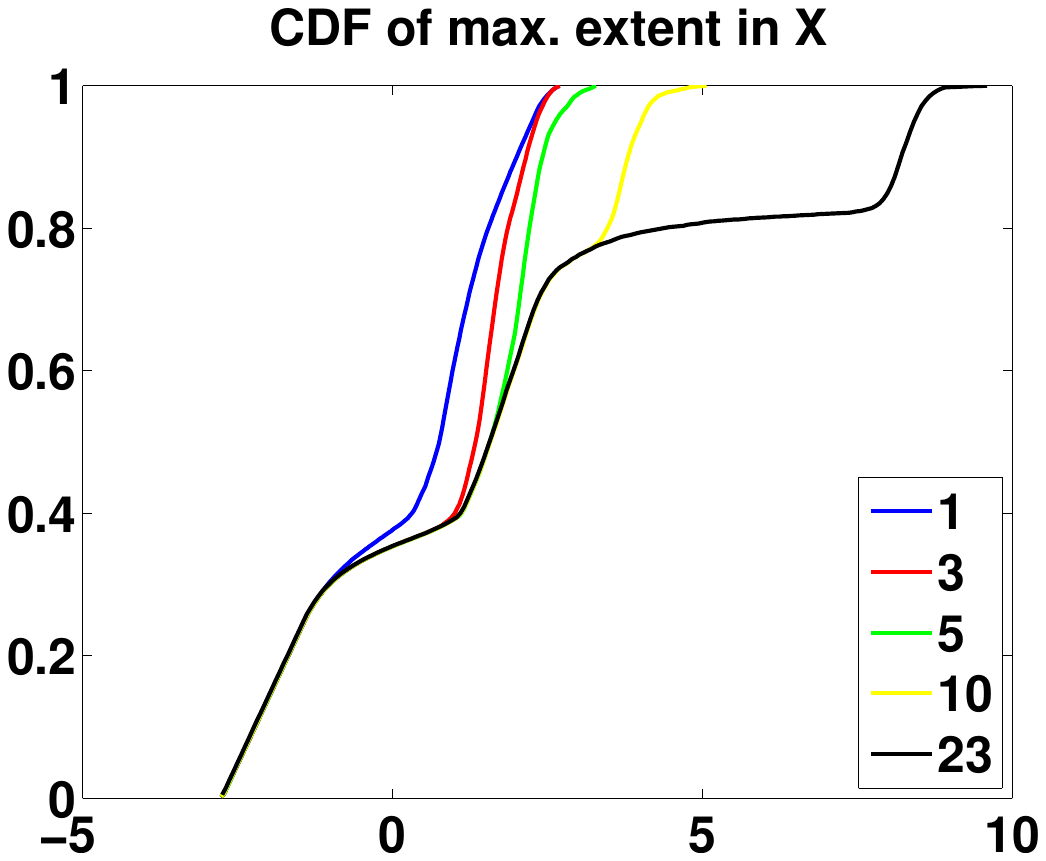}
  \includegraphics[width=7cm]{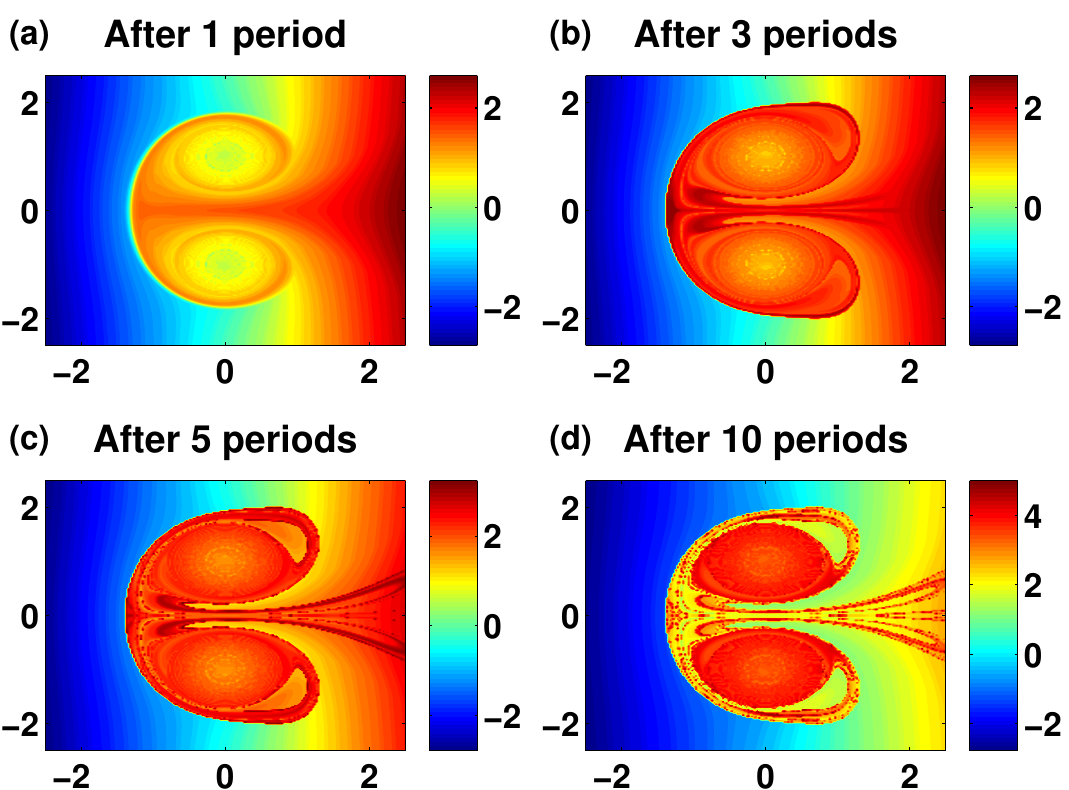}
  \caption{The MET dependence on frame of reference. The OVP flow with
    the same parameters as in Fig.~\ref{fig:movingtime} yet in a
    different moving frame (Eq.~\ref{eq:psivortexpair} with
    \(\epsilon=0.2,\omega=1.45\) and \(v=0.5\) ). Comparing the right
    panels here with those of Fig.~\ref{fig:movingtime} we observe
    that similar structures appear and the main change is in the color
    bar. Indeed, the left panel showing the CDF of the slower moving
    frame shows that after the initial transition the frame of
    reference only shifts the location of the CS, as expected.  }
\label{fig:secondframe}
\end{figure}

Fig.~\ref{fig:movingtime} shows the maximal extent in \(x\) and the
associated CDFs of the OVP flow at several extremal windows of
increasing length. The CSs appear as a sharp parabolic increase in the
CDF, its center moving with time. Thus it is possible to detect the
area and location of the body of fluid which advects with the vortices
as well as to identify the volume of fluid that is shed by the lobes.

Fig.~\ref{fig:secondframe} shows, as in Fig.~\ref{fig:movingtime}, the
maximal extent in \(x\) of the OVP flow, but in a different moving
frame. After some time, aside of the change in the color bar, similar
structures as in Fig.~\ref{fig:movingtime} appear, as is also apparent
from the CDFs of the two simulations. Thus, it is demonstrated that
for sufficiently large \(\tau\) the MET is frame-independent
\cite{PoHa99,hallerjfm13} and applicable even when the structure moves
out of the original region in which the particles are seeded. Note
that these features are especially relevant to geophysical
applications in which the observed domain is open and there is an
unknown underlying current, see section \ref{sec:realdata}.

\begin{figure}
\centering
  \includegraphics[width=5cm]{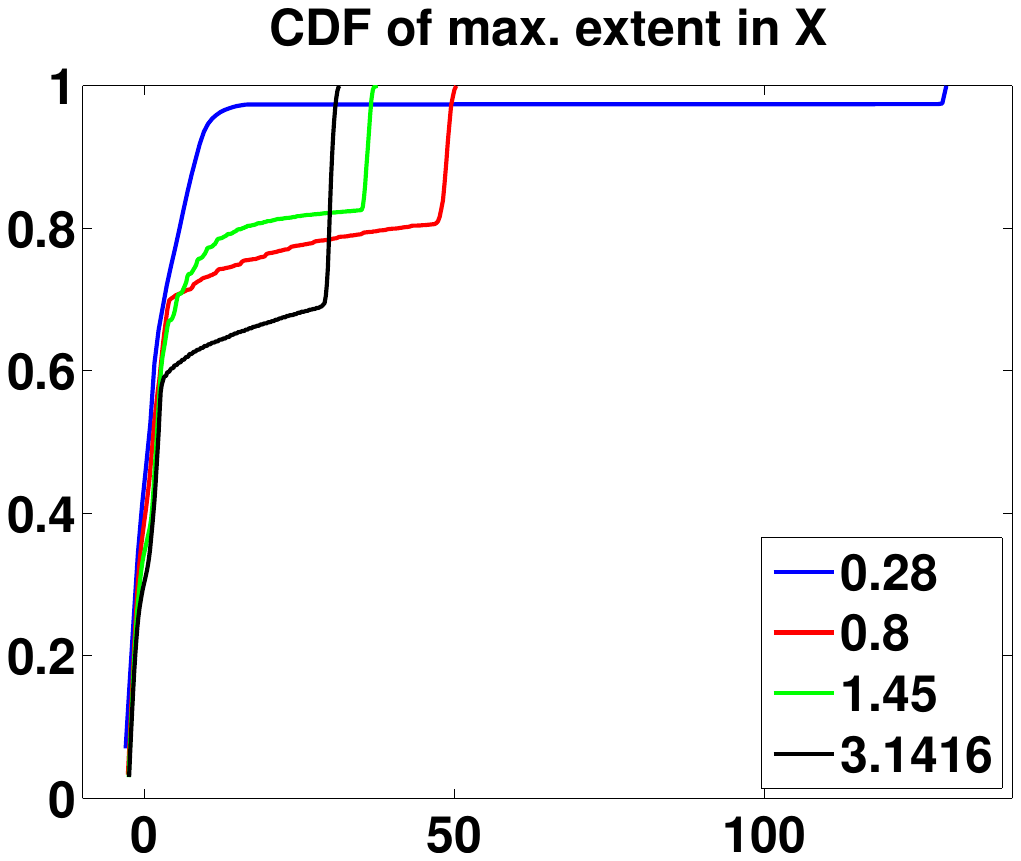}
  \includegraphics[width=7cm]{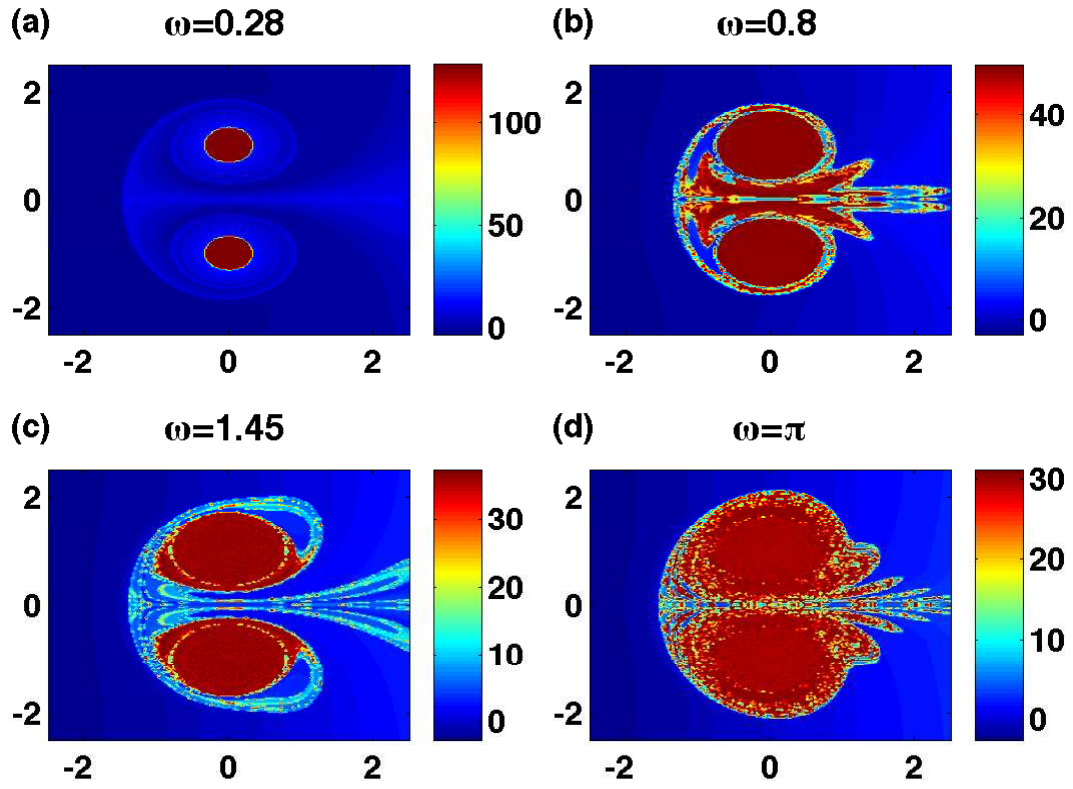}
  \caption{ CSs dependence on parameters. On the right panels the
    maximal extent in \(x\), $M^{+}_{\bf (1,0)}$, is shown for four
    different frequencies at $\tau=100$ (about 5, 12, 23 and 50 cycles
    of the corresponding frequencies). The left panel shows the CDF of
    these fields - the area and location of the CSs is easily
    extracted from the CDFs. Notice the non monotonic dependence of
    the core area on $\omega$.}
\label{fig:3omegas}
\end{figure}

  \begin{figure}
\centering
\includegraphics[width=5cm]{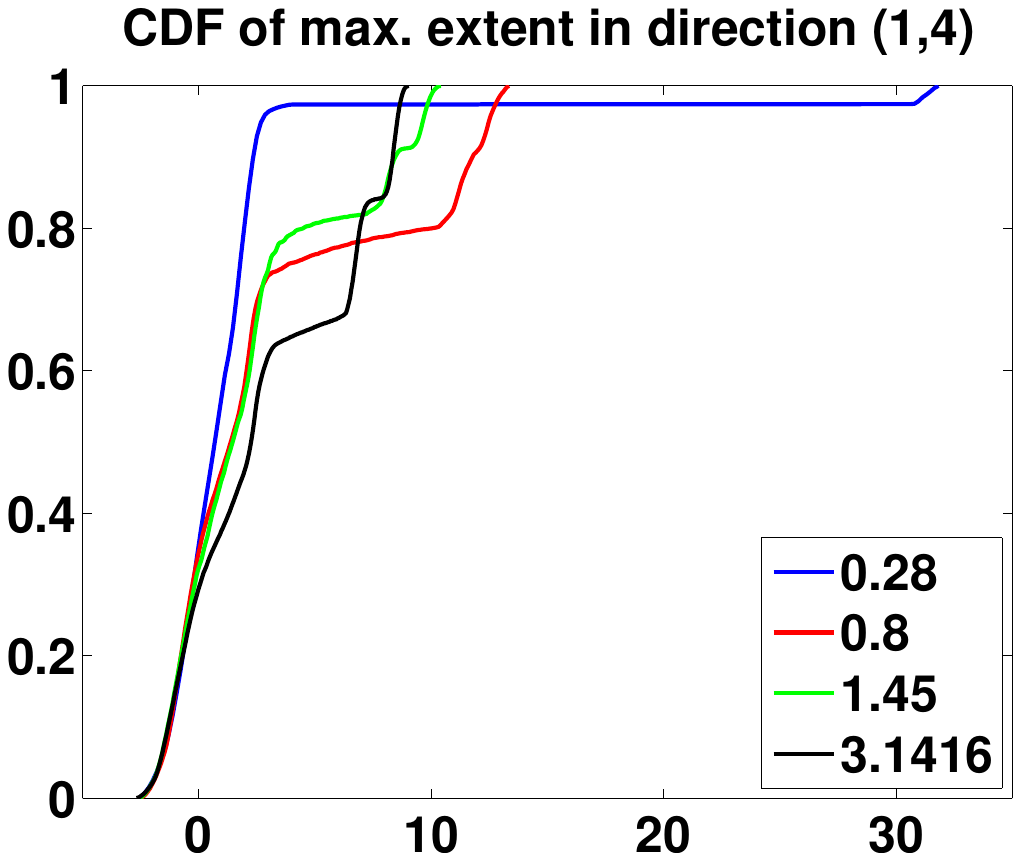}
\includegraphics[width=7cm]{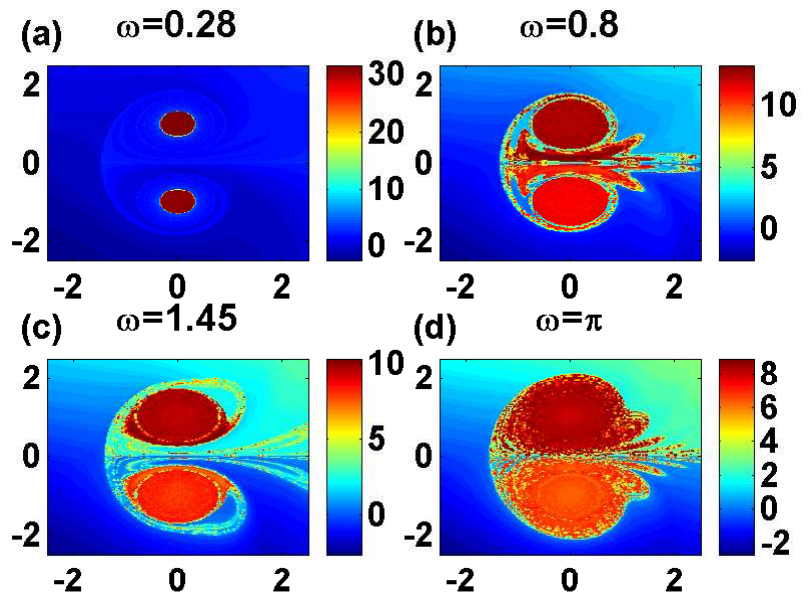}
    \caption{CSs in a resolving direction. On the right panel the
      maximal extent in \(x+4y\), \( M^{+}_{\bf (1,4)}\), is shown for
      four different frequencies (as in Fig.~\ref{fig:3omegas}). The
      left panel shows the CDF of these fields. Here, the resolving
      direction distinguishes between the lower and upper vortex
      areas.}
\label{fig:xplus4y}
\end{figure}

Fig.~\ref{fig:3omegas} (respectively \ref{fig:xplus4y}) shows the
maximal extent in \(x\) (respectively in direction {\bf r}=\((1,4)\))
of the OVP flow for different frequencies of the strain field
oscillations. The CSs area has a strong non monotonic dependence on
the frequency (see also \cite{RkPo99}), as is apparent from both the
CDF diagram and the extremal field plots.  While the \(x \) direction
lumps together both vortices, the \((1,4)\) direction resolves the two
structures.

\begin{figure}
\centering
\includegraphics[width=7cm]{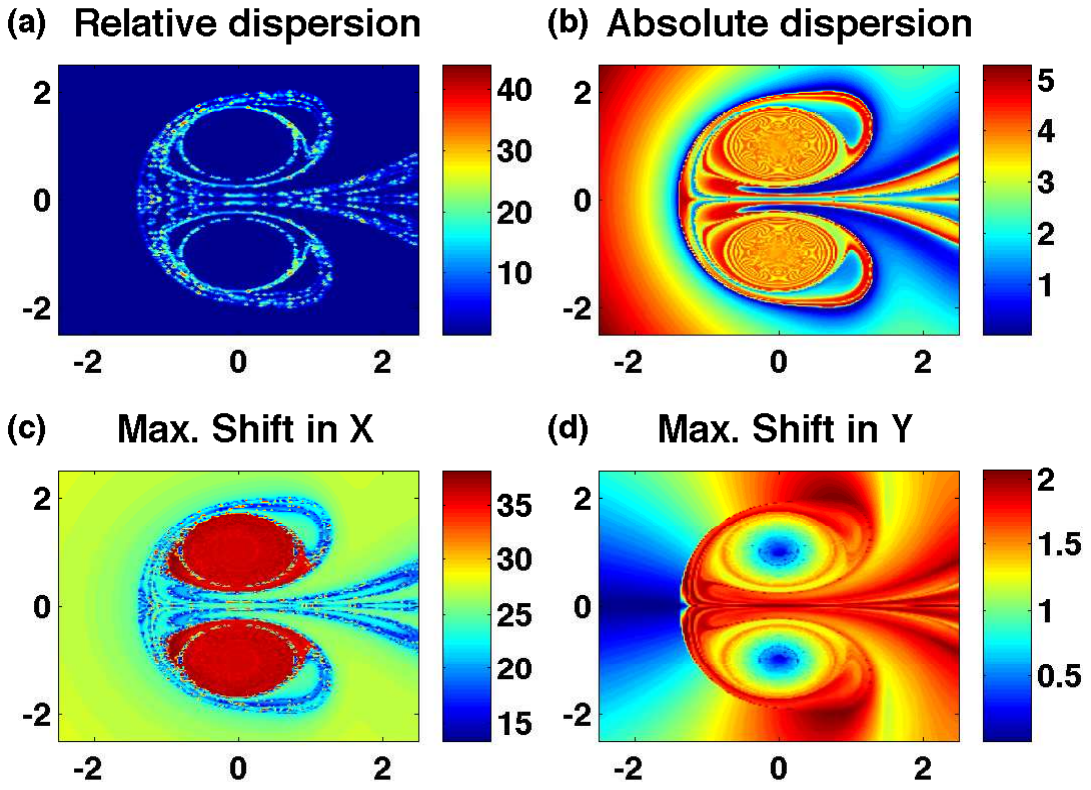}
\caption{ Different quantifiers for the OVP flow.  a) The relative
  dispersion. b) The absolute dispersion c) The maximal shift in
  \(x\), \(M_{\bf(1,0)}\), d) The maximal shift in \(y\),
  \(M_{\bf(0,1)}\).  Compare with Figs
  \ref{fig:3omegas},\ref{fig:xplus4y}c in which the same flow is
  presented with the max. extent plots (Eq.~\ref{eq:psivortexpair}
  with \(\epsilon=0.2,\omega=1.45,\) \(v=0.25 \) at \(\tau=100\),
  i.e. after about 23 cycles).  }
\label{fig:shiftadrd}
\end{figure}

Finally, Fig.~\ref{fig:shiftadrd} shows several other quantifiers to
be compared with the maximal extent in \(x\). Fig.~\ref{fig:shiftadrd}
a (b) presents the traditional relative (absolute) dispersion fields
and Fig.~\ref{fig:shiftadrd}c (d) presents the maximal shift in \(x\)
(\(y\)). Since some of the particles move to the right (the vortex
core region) and some to the left (the outer particles), the maximal
shift in \(x\) field together with the maximal extent field provide
information on the directional motion of the particles. The maximal
shift in \(y\) (d) and the absolute dispersion (AD) field (b) provide
similar division to core regions, mixing region and outer flow, yet
these do not include directional information. The RD field (and
similarly the FTLE) nicely detects the stable manifold, yet has
similar, close to zero values everywhere else.

It is worth pointing out the differences between moving CS and CS that
has on average a zero displacement. To gain intuition, again assume
that trajectories belonging to a CS which moves in an unbounded region
along a direction \(\textbf{r}\) have the form
\(x(t;t_{0})=x_{c}(t;t_{0})+g(t;x_{0}) \) where \(x_{c}(t;t_{0})\) is
the CS center, so \(x_{c}(t;t_{0})\cdot\mathbf{r}\) is slowly
increasing on average, and the rapidly oscillating \(g \) has average
zero. Then, while \(M_{\mathbf{r}},M_\mathbf{r}^+\) may be unbounded
in time, \(M_\mathbf{r}\) and \(M_{\mathbf{r}}^+\) still attain their
local minima at \(x_{c} \), see
Figs.~\ref{fig:movingtime}-\ref{fig:shiftadrd}.  On the other hand, in
this case \(M_{\mathbf{r}}^-\) contains very little information with
regard to the CS - basically the launching point of the trajectories.
If one chooses a direction which is perpendicular to the direction of
motion (i.e. \(y\) direction here), the number of separate regions may
be detected, but the information on the CS motion is lost.

Trajectories that are far from the vortices experience, on average, a
nearly uniform flow, hence the MET have a nice smooth dependence on
initial conditions in such regions. The CDFs thus have a bulk smooth
region corresponding to the background flow, a quadratically
increasing portion with a moving center that corresponds to the CS,
and some shedding of lobes that appear as small steps in the CDF, see
Figs.~\ref{fig:movingtime}-\ref{fig:3omegas}. The speed and size of
the moving CS can be thus easily determined from the CDF.  Notice that
the CS area is the relative fraction obtained from the CDF multiplied
by the chosen seeding area (see below).

\section{\label{sec:realdata}Real data}

\begin{figure}
\centering
\centering
  \includegraphics[width=9cm]{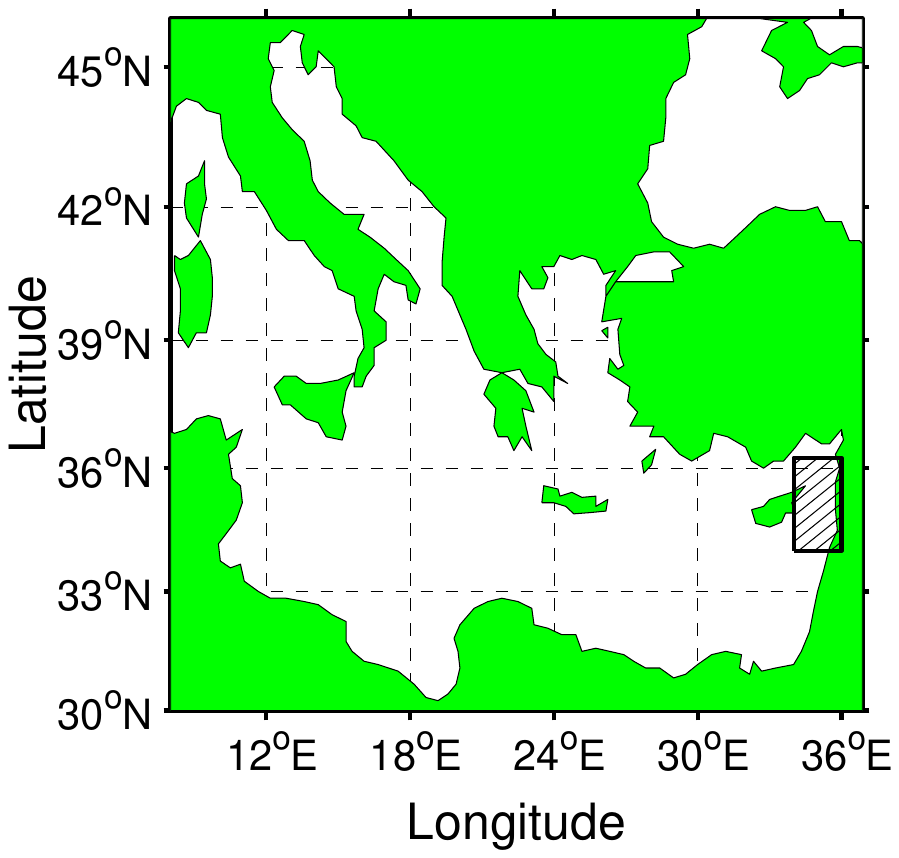}
  \caption{The eastern Mediterranean sea region in which particles are
    seeded (dashed rectangle) is contained in a large region where
    velocity data is available.}
\label{fig:realdataregion}
\end{figure}

We next apply the MET analysis to real data from the eastern
Mediterranean, using surface currents obtained from the AVISO database
(http://www.aviso.oceanobs.com). Within this velocity field we deploy
10,000 virtual particles on an approximately 2 km grid and track them
for 40 consecutive days. The particles are seeded in a much smaller
domain than the domain covered by the altimeter so that even when they
leave the initial seeding region their trajectories can still be
computed, see Fig.~\ref{fig:realdataregion} (transport properties in
this region during this period were studied in \cite{Efrati13}).
During the period examined in this study, the distributed global
product was a combination of altimetric data from Jason-1 and -2 and
Envisat missions. The dataset is comprised of daily near-real-time
sea-level-anomaly data files, gridded on a $\frac{1}{8}^o \times
\frac{1}{8}^o$ Mercator grid. The methodology for extracting a
velocity field from sea level data is known to introduce errors, as
does the linear interpolation scheme we use for integrating
trajectories.  Other sources of uncertainties in the data are due to
tides and atmospheric conditions. In particular, the extracted
velocity field is not area-preserving.  Although in stratified ocean
the flow is approximately 2D (i.e. vertical velocities are few orders
of magnitude smaller than the horizontal velocities), 3D effects may
qualitatively changes surface mixing \cite{AhGilRK12}.  Additionally,
close to the coast, the use of satellite altimetry is known to be
unreliable, so the dataset does not include measurements at distances
of less than 10 km from the coastline. Despite these above-mentioned
errors and limitations, our analysis seems to capture the existing
CSs.

\begin{figure}
\centering
  \includegraphics[width=5cm]{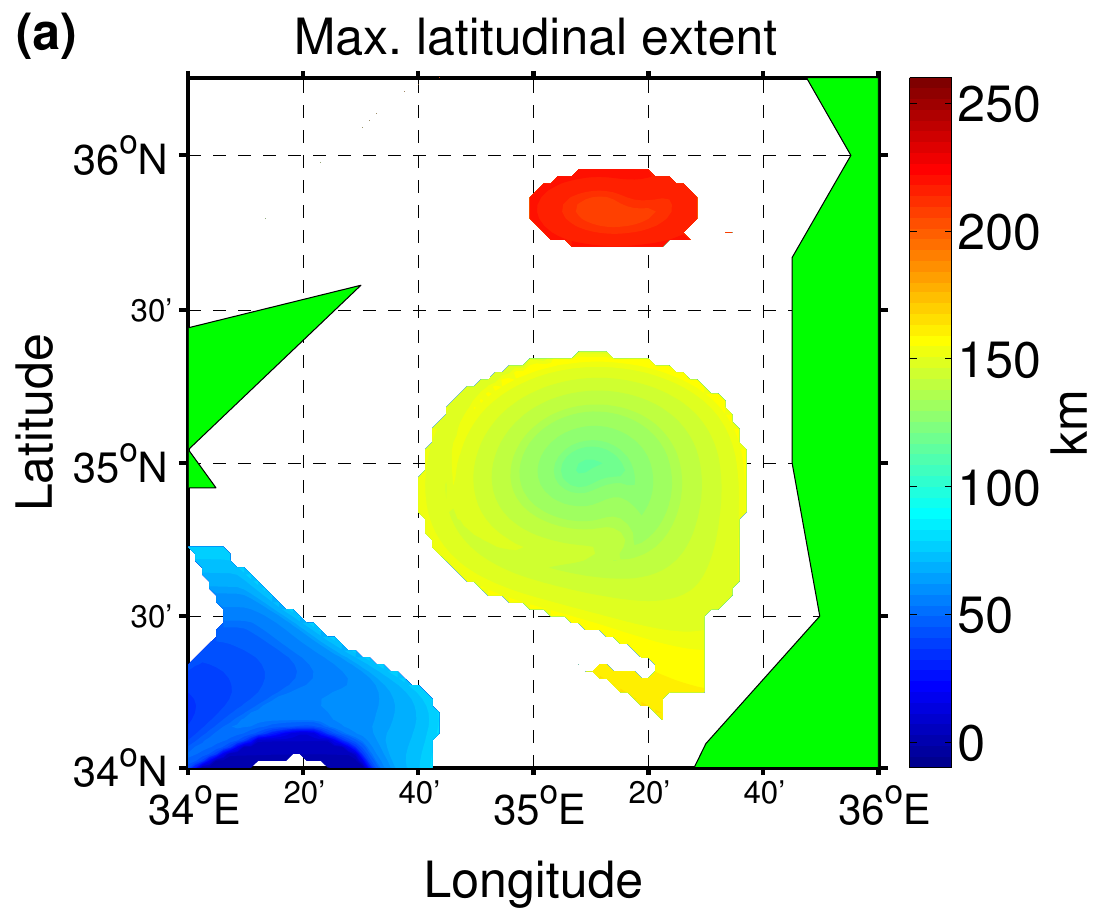}
  \includegraphics[width=5cm]{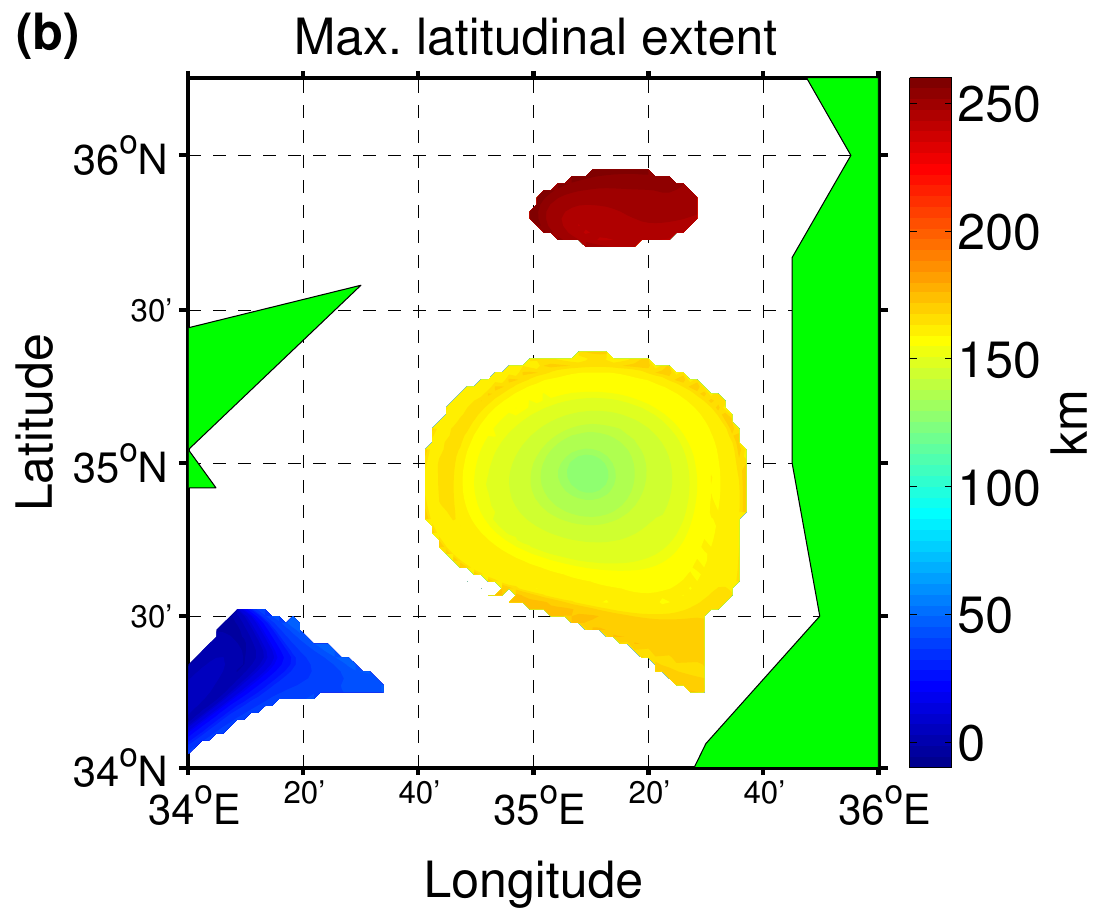}
  \includegraphics[width=5.5cm]{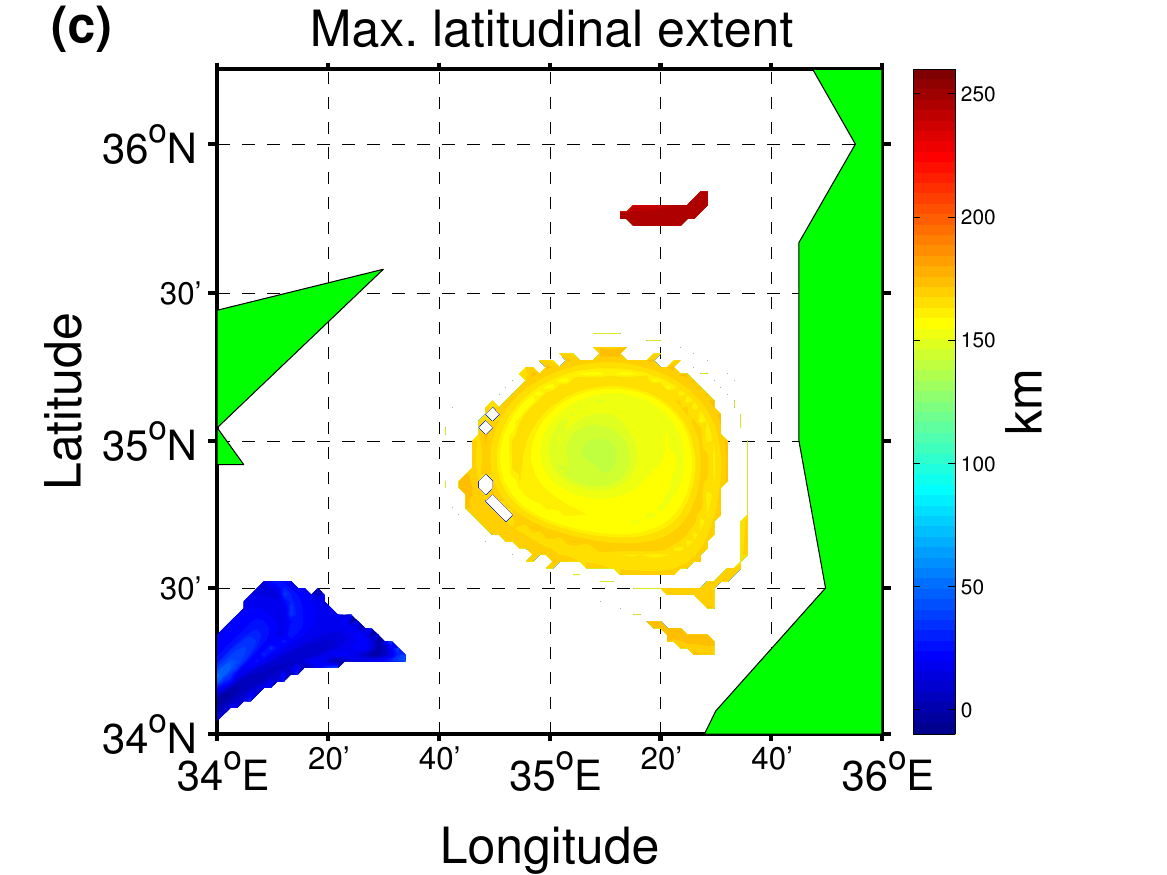}
  \caption{The maximal latitudinal extent (\(M^+_{\bf (0,1)}\)) of
    particles seeded on 18 of Aug. 2011 in the eastern Mediterranean
    sea. Three separate CSs are identified.  Extremal windows of 10
    days each are shown a) Aug. 18-Aug. 28 b) Sep. 6-Sep. 16 c) Sep.
    16-Sep. 26. White regions correspond to trajectories which either
    originated or reached domains with no reliable velocity data
    (either approached the coastal area or left the region depicted in
    Fig. \ref{fig:realdataregion}).  }
\label{fig:realdata}
\end{figure}

\begin{figure}
\centering
 \includegraphics[width=5cm]{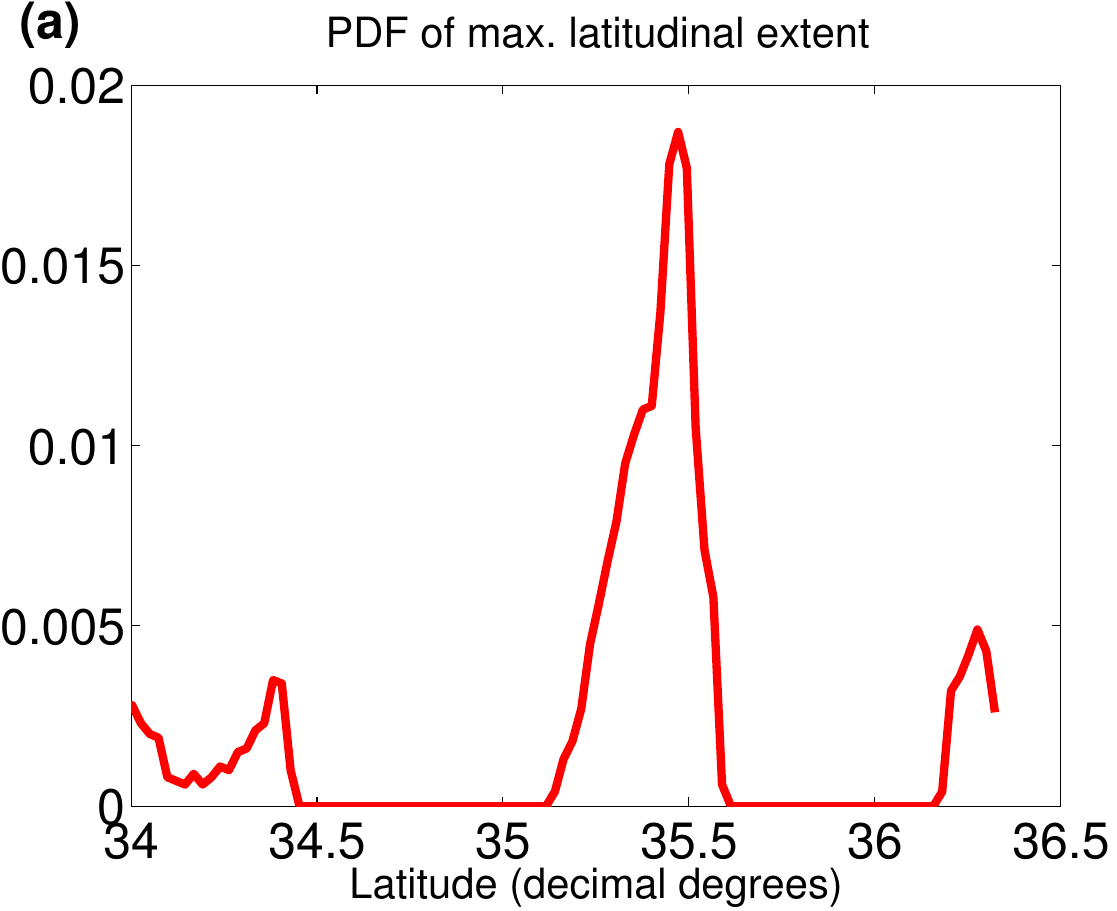}
 \includegraphics[width=5cm]{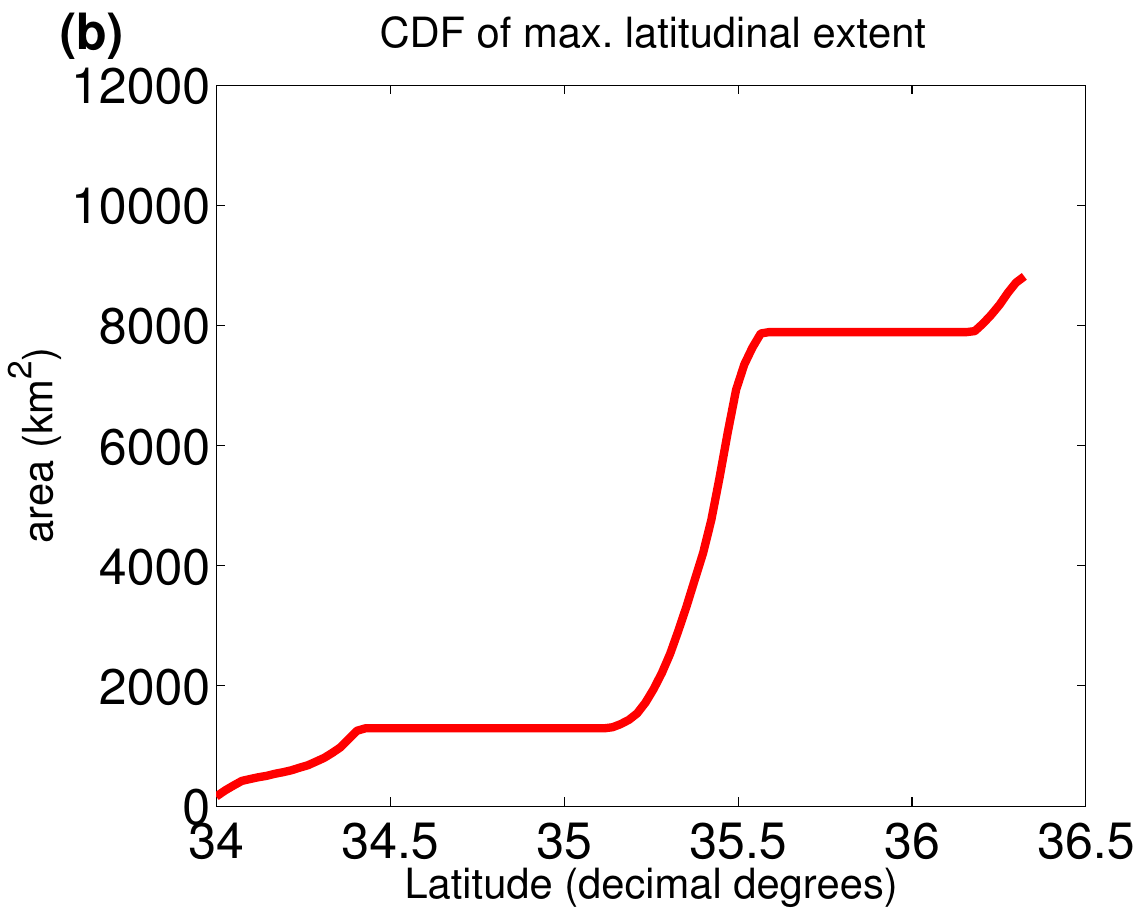}
  \includegraphics[width=5cm]{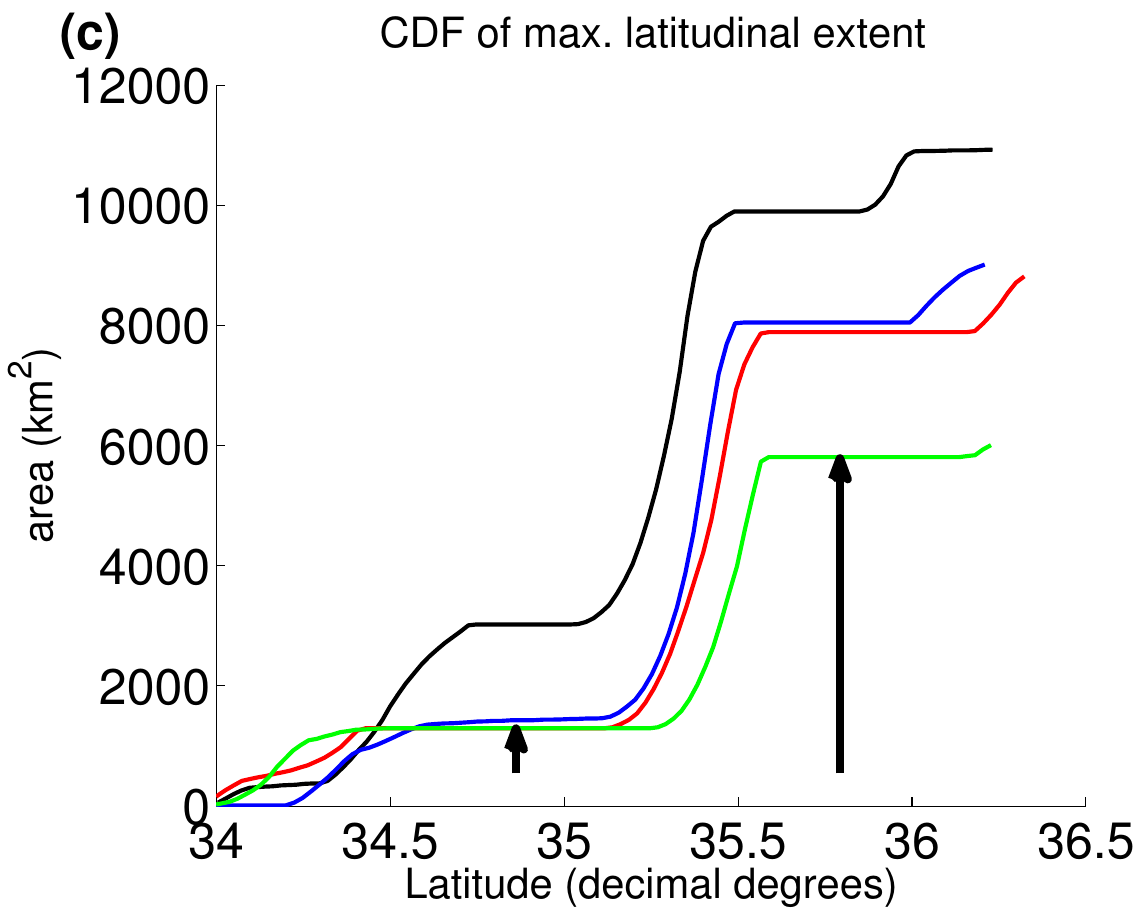}
  \caption{The PDF (a) and CDF (b) of the maximal latitudinal extent
    (\(M^+_{\bf (0,1)}\)) of particles seeded on 18 of Aug. 2011 for
    the extremal window Sep. 6-Sep. 16 (Fig \ref{fig:realdata}b).  The
    three distinct regions are clearly seen in the figures. The
    signature of the two CS which are fully contained in the region
    are similar to those appearing in the toy models. (c) shows the
    CDF for four 10-days periods: Aug. 18-28, Aug. 28-Sep. 6,
    Sep. 6-16, Sep. 16-26. Notice the changes in the CSs centers'
    latitudes, the disappearance of the upper gyre and the dramatic
    decrease in the main gyre area - the gap between the two plateaus indicated by the arrows. }
\label{fig:realdatapdf}
\end{figure}

\begin{figure}
\centering
  \includegraphics[width=4cm]{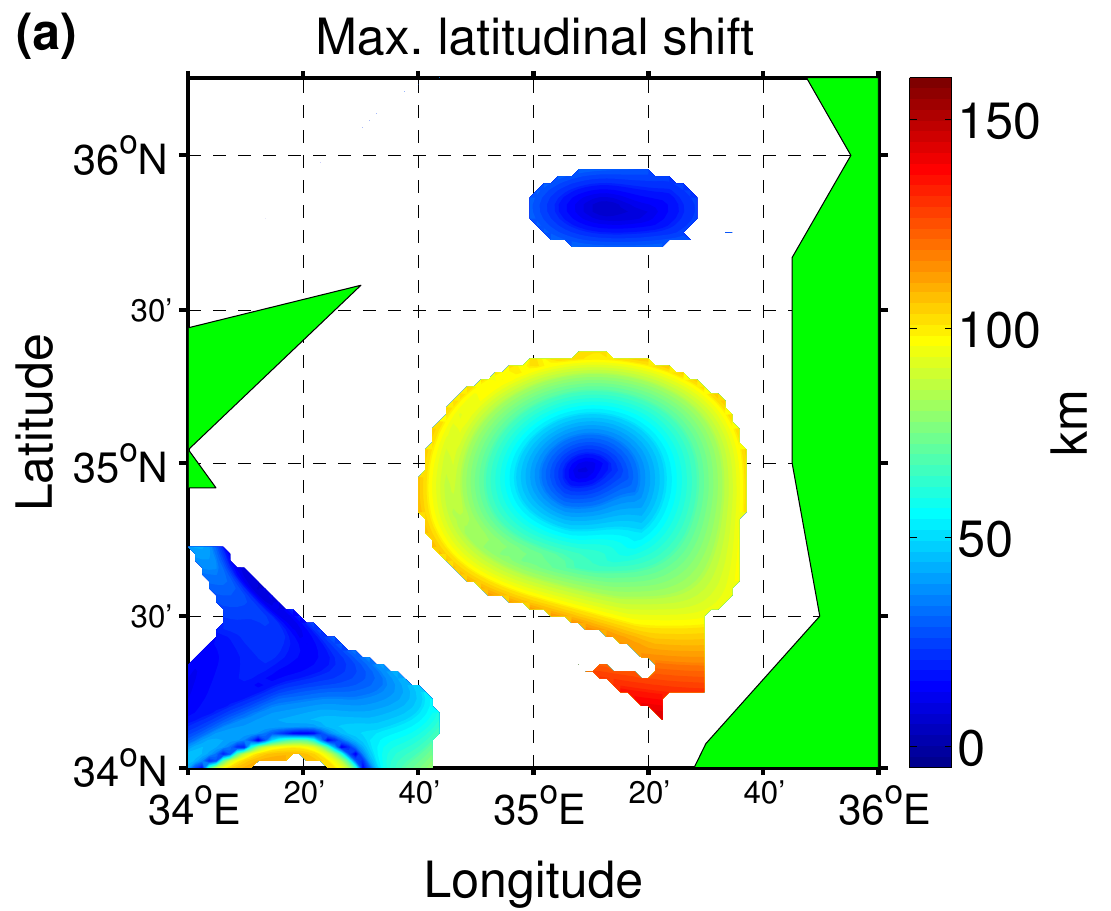}
  \includegraphics[width=4cm]{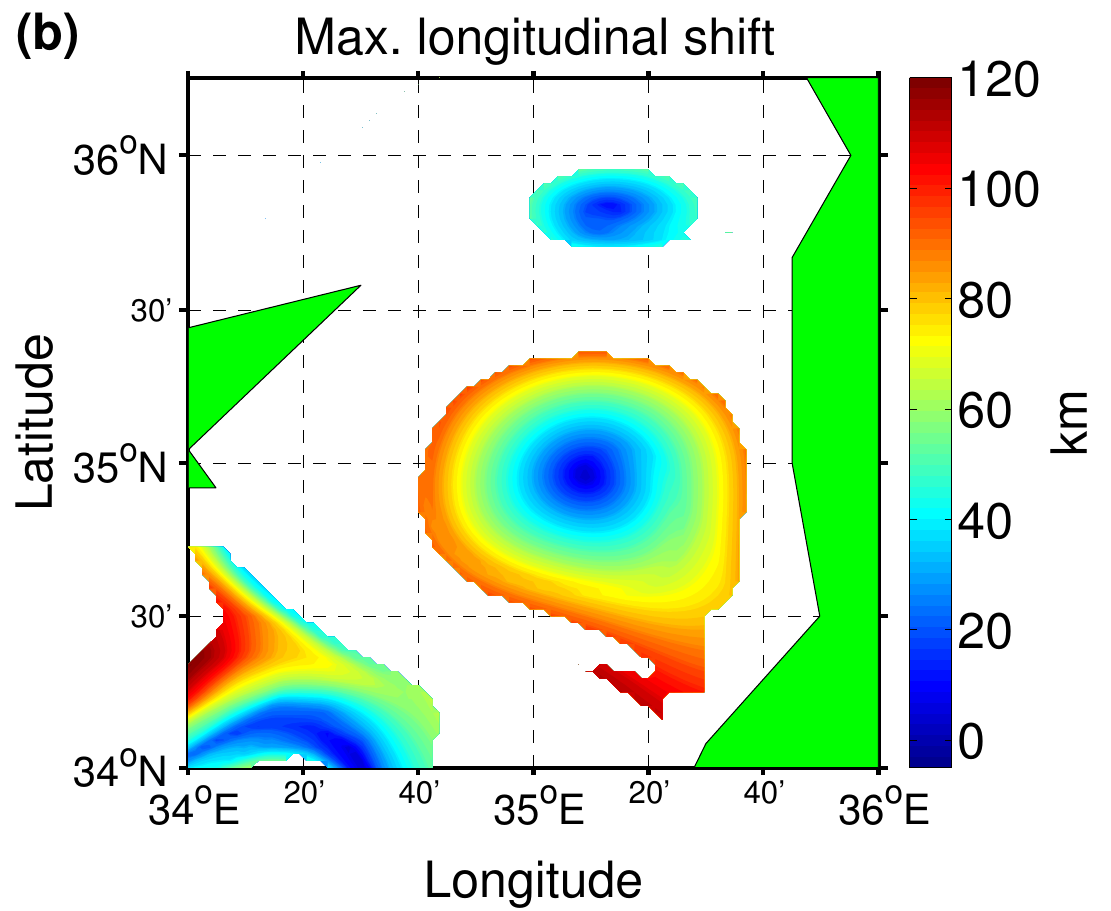}
  \includegraphics[width=4cm]{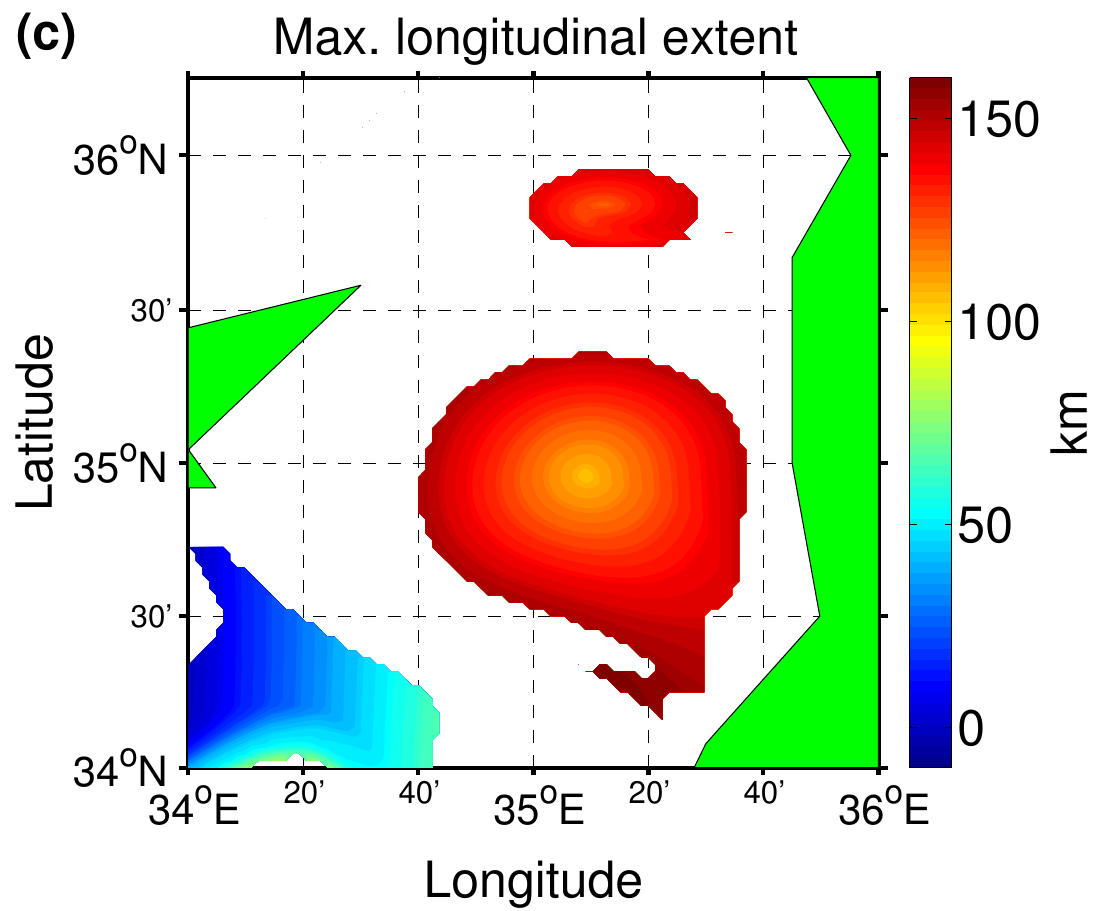}
 \\
  \includegraphics[width=4cm]{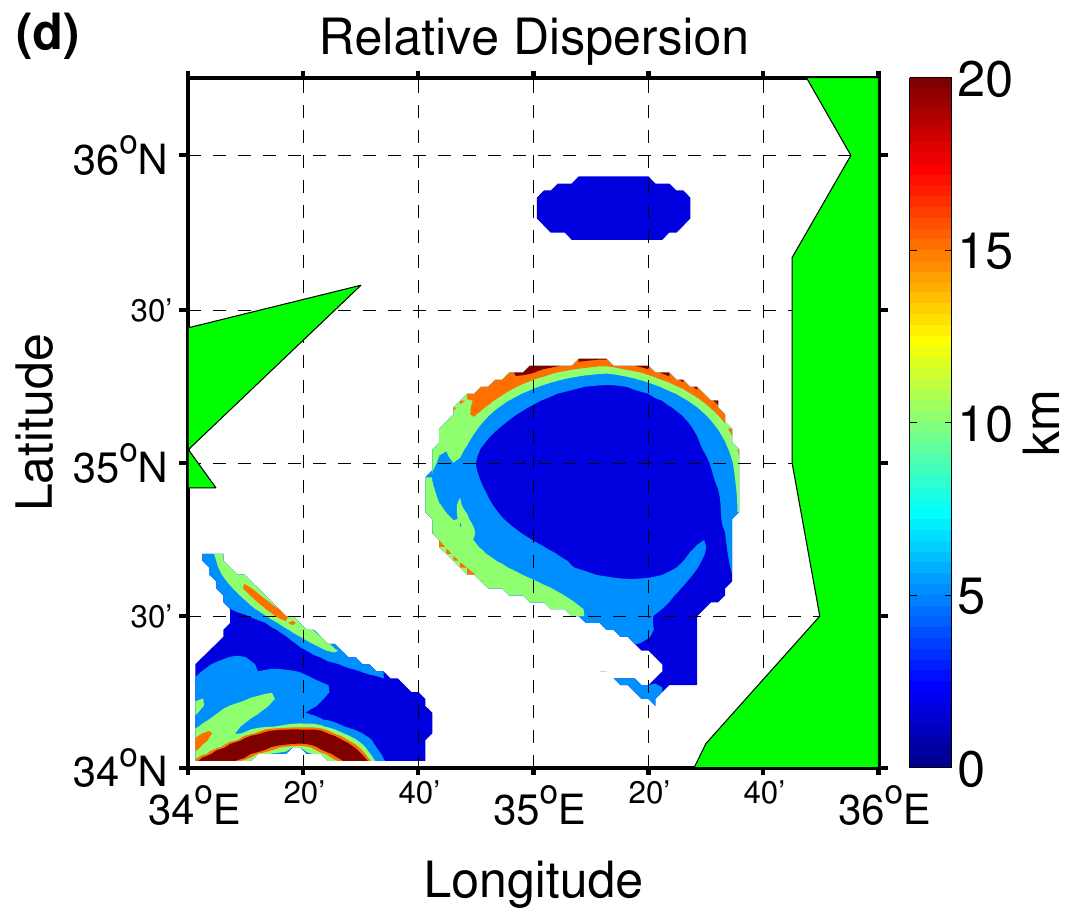}
  \includegraphics[width=4.1cm]{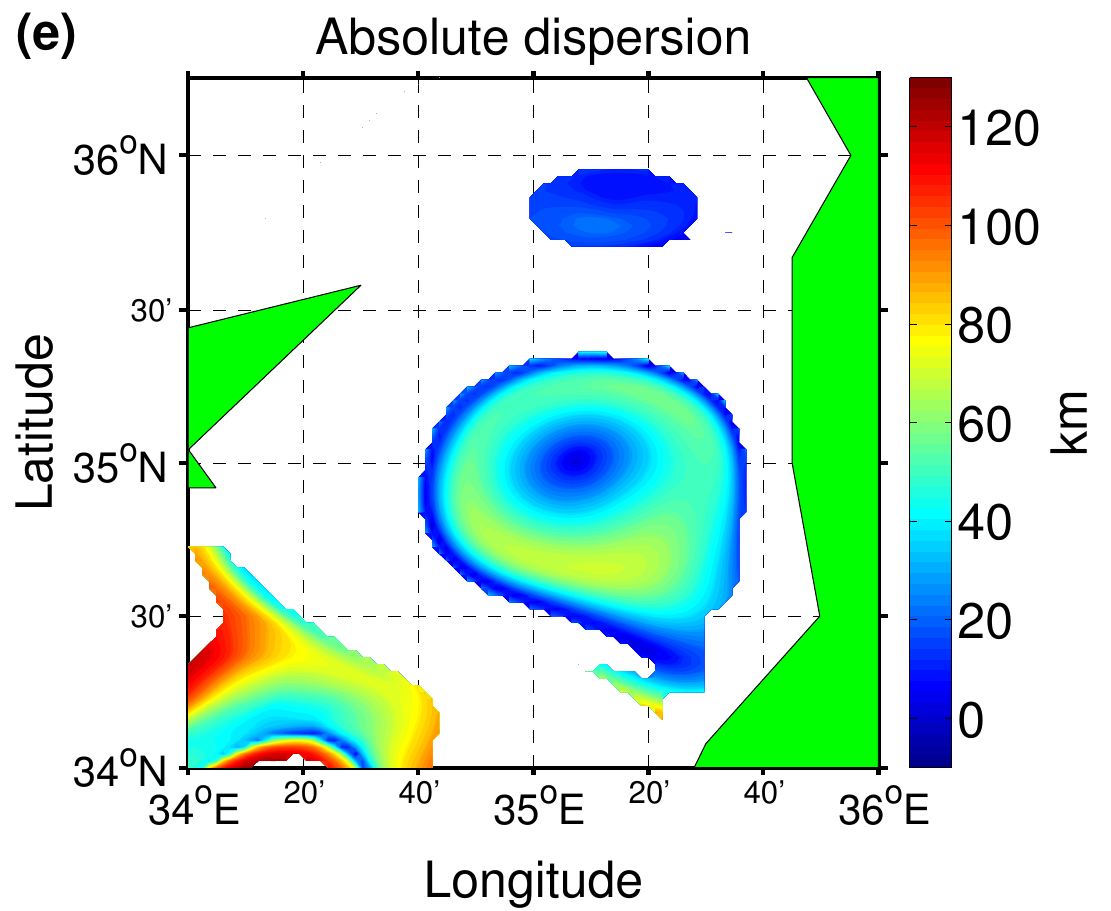}
  \caption{ Other measures for the particles seeded on 18 of Aug.
    2011 on the extremal window Aug. 18-Aug. 28. a) The maximal
    latitudinal shift (\(M_{\bf (0,1)}\)) b) The maximal longitudinal
    shift (\(M_{\bf (1,0)}\)) c) The maximal longitudinal extent d)
    the absolute dispersion e) the relative dispersion. }
\label{fig:realdatamad}
\end{figure}

Since we do not have apriori knowledge of the flow field, we
calculated both minimal and maximal extents along both the zonal
(longitudinal) and meridional (latitudinal) directions for a few
extremal time windows. The maximal extent in the latitudinal direction
provided the best separation between the two complete CSs (centered at
$35^o$ and $35^{o}50^{'}$ north) that were detected in essentially all
measures (see Figs.~\ref{fig:realdata}-\ref{fig:realdatamad}).  The
white area in these plots corresponds to trajectories that either
originated or reached domains with no reliable velocity data by the
end of the extremal window integration time (either approached the
coastal area or left the region depicted in
Fig. \ref{fig:realdataregion}).

Fig.~\ref{fig:realdata} shows the maximal extent in the latitudinal
direction for three 10 days extremal time windows starting on Aug. 18,
Sep. 6, and Sep. 16 (the Aug. 28 panel, not shown, is very similar to
that of Sep. 6).

Fig.~\ref{fig:realdatapdf}a,b shows the PDF and CDF of the maximal
extent in the latitudinal direction for the middle extremal time
window (Sep.  6-16). The CDF scale is multiplied by the approximate seeding area to obtain realistic information regarding the gyres size. Three distinct structures are clearly seen. The
two CSs that are fully contained in the region (centered at $35^o$ and
$35^{o}50^{'}$ north) are manifested in the PDF and CDF with the
typical asymmetric form, similar to
Figs.~\ref{fig:pdfcdf}-\ref{fig:oscilcslong}. On the other hand the
structure that is only partially contained in the domain in its
south-west corner cannot be identified as a CS in the CDF. This
suggests that we are seeing only a part of the CS or of another
structure - to be resolved, a larger domain is needed.  Fig
~\ref{fig:realdatapdf}c shows the CDF of the maximal extent in the
latitudinal direction for four subsequent 10 days extremal time
windows.

From the CDF plots it is seen that the main CS, centered near latitude
\(35^{o}\), keeps its latitude and size for the first 30
days, and then, in the last 10 days it shifts a bit north-word and a
substantial part of its outer layer approaches the coastal area. The
smaller northern CS, centered originally near latitude $35^{o}50^{'}$
north, is seen to travel north-word right from the start (e.g. notice its color change between Fig.~\ref{fig:realdata}a and
Fig.~\ref{fig:realdata}b ) and in the last 10 days most of the
trajectories in it approached the coastline.

Fig.~\ref{fig:realdatamad} shows a few more fields that provide
additional information on the CS structure and demonstrate the
importance of the choice of a resolving
direction. Fig.~\ref{fig:realdatamad}a shows the maximal latitudinal
shift, Fig.~\ref{fig:realdatamad}b,c show the maximal longitudinal
shift and extent and Fig.~\ref{fig:realdatamad}d,e show the standard
absolute and relative dispersion fields. In all these figures the two
complete CSs are nicely seen. However, since the two gyres have
overlapping field values (have the same color in the colormap), a PDF
and CDF of any of the five presented fields will not exhibit the
separate structures as seen in Fig.~\ref{fig:realdatapdf} for the
resolving direction (here latitude). Of course, one can use a spatial
fraction of the domain to isolate each of the gyres, and these
specific plots also provide additional information regarding the
longitudinal motion. Notice also that the RD field is more expensive
computationally than any of the MET characteristics, and its
computation in different time windows requires reseeding of particles
(whereas all other computations are done by processing a single
40-days integration of the trajectories).

% \begin{figure}
% \centering
% % \\
% % hezi \bigskip \\
% %
% \caption{   }
% \label{}
% \end{figure}

\section{Discussion}

Our main result is the introduction of a new family of Lagrangian
diagnostics, in particular the MET, and the demonstration that the CDF
of the MET allows one to find important characteristics of the CSs at
low computational cost. In particular, the number, location, and size
of the coherent structures (nested sets of a continuum of ergodic
components) with a volume above a threshold value may be found with no
need for minimization or image processing procedures. A major
advantage is thus the ability to compress a large amount of data into
a simple diagnostic plot.

The signature of a CS in the CDF appears as a smooth curved increasing
segment - the base of it and the value where it flattens or abruptly
increases provide information on its spatial location and width along
the specific direction that is used to compute the MET. The height
difference between these values provides its area. The signature of a
mixing layer (in closed domains) is a fast growing segment that
asymptotes as \(\tau\), the extremal window, grows to infinity, to a
discontinuity of the CDF.  A motion of the CS corresponds to a shift
of its corresponding segment in the CDF with hardly any change of its
shape. We demonstrated that the MET provides insightful information
using toy models in both closed and open domains and on a real data set
from the eastern Mediterranean Sea.

The MET and many other Lagrangian quantifiers (such as RD, FTLE, the
hypergraph map and other averaged quantifiers
\cite{budivsic2012applied,Mezic10,Ryp12,bozorgmagham2013real,Shadden05,Haller00})
have a common feature: asymptotically these converge to constants on
ergodic sets and hence, in principle, may be utilized to divide the
phase space to separate ergodic components. In many applications, the
transient properties of these and other Lagrangian quantifiers were
studied, showing that in some cases ridges of finite time realizations
of these fields provide good predictors for dividing surfaces. We
expect that similar analysis can also be applied to finite time
realizations of extremal fields (see especially
Figs.~\ref{fig:oscilcs} and \ref{fig:movingtime}) and this direction
has yet to be explored. Here we exploit the asymptotic features of
these fields as a way to identify CS. In this aspect, we note that the
RD and FTLE are degenerate - they asymptote to zero in the regular
regime and to a positive constant in the mixing zone. On the other
hand, the value of the MET (and of the hypergraph map and other
Lagrangian averages \cite{Mezic10}) asymptotes to a smooth function in
a regular region and to a constant in the chaotic zone. The unique
feature of the MET is that if the direction \textbf{r} is resolving,
its CDF readily provides additional information on the location and
number of CSs (whereas with the other quantifiers the CDF lumps
together all coherent structures).

Another distinguishing characteristic of the MET is that the
convergence to its asymptotic value is always monotone in time whereas
in all the other quantifiers convergence to their asymptotic values is
oscillatory (except the arc-length map
\cite{MendozaManco10,Alvaroet1013} which is monotone yet
unbounded). Hence we expect that the convergence of the MET will be
faster and more regular. In fact, the temporal convergence properties
of the MET may be related to the universal convergence associated with
extreme value statistics of ergodic dynamical systems
\cite{Collet2001,Holland2012,Lucarini2012,Freitas2008,Freitas2013}. The
implications of these on the convergence of the CDF, on the spatial
smoothness of the MET, and on the sensitivity of these to noise and
velocity errors have yet to be investigated.

The current work leaves many additional directions to be explored in
future studies, including: (1) The transient MET fields in the
coherent structures are quite smooth whereas their transient behavior
in the mixing layers is noisy. This property may be utilized to
distinguish between these regions on short time scales. More
generally, the study of the transient behaviour of the MET in \(\tau ,
t_{1}\) may reveal the structure (e.g. local dimension
\cite{Lucarini2012,Ryp12}) of the ergodic component. Possibly, it may
reveal other transient transport processes, such as dividing surfaces
(LCS) and the lobe structure \cite{thesis08}.  (2) The applications
shown here suggest that additional insight may be obtained by studying
the temporal and spatial dependence of extreme values of other
observable functions in systems with mixed phase space
(e.g. velocities, speed, distance from the origin, stretching rates,
strain rates, FTLE, recurrences \cite{Lucarini2012} etc.)  (3) The
study of open flows needs to be further explored. For open flows with
moving CSs the difference between the traditional AD field and the MET
fields is not dramatic, yet the MET fields provide the additional
advantage of directional information and non-oscillatory
convergence. (4) In real applications we often have a limited sampling
of the flow along specific directions, e.g. by few drifters. How much
of the flow characteristics can be extracted from such limited
information is unclear. Based on our preliminary results (not shown),
 1D sections might suffice to provide many of the flow
characteristics. Moreover, this approach may work in higher dimensions
as well. (5) In flows with a large number of CSs of different scales
and locations, such as 2D-turbulent flows, it may become difficult to
find resolving directions. In such cases it may be beneficial to adopt
a multi scale strategy by which resolving directions are sought on
subdomains. (6) Finally, we note that many of the above issues may be
studied first on chaotic maps with mixed phase space behavior (e.g. we
currently study the standard map and its higher dimensional
extensions). Such studies allow the introduction of a more rigorous
mathematical analysis.

In conclusion, we present new, promising Lagrangian diagnostics that
enable the extraction of properties of coherent structures from large
data sets by looking at extremal values of observables, their PDFs,
and CDFs. These diagnostics are not only simple, intuitive, and
computationally cheap; they also enable a significant data reduction, since it is possible to extract from the cumulative distribution functions much of the relevant information regarding the existence, location, size and motion of the coherent structures.

\begin{acknowledgments}
  We thank Dimitry Turaev for useful comments and suggestions. This
  study was supported by grants from the Israel Science Foundation (to VRK and to EF) and
  MOST (to HG). This work was stimulated from the MSc thesis of Rotem
  Aharon \cite{thesis08,AhGilRK12}.
\end{acknowledgments}

%\bibliography{dgyre_refs,fullref} % Produces the bibliography via
                                % BibTeX.
%\bibliography{dgyre_refs} % Produces the bibliography via BibTeX.

\end{document}